\documentclass{aa}  
\usepackage{graphicx}
\usepackage{orcidlink}
\usepackage{txfonts}
\usepackage{xcolor} 
\usepackage{natbib}
\bibpunct{(}{)}{,}{a}{}{,}
\usepackage{hyperref}
\hypersetup{
    colorlinks,
    citecolor=cobalt,
    linkcolor=[rgb]{0.8, 0.2, 1.0},
    urlcolor=cobalt,
    breaklinks=true
}

\definecolor{cobalt}{rgb}{0.06, 0.2, 0.65}

\newcommand{\Eq}[1]{Eq.~(\ref{#1})}          
\newcommand{\Equation}[1]{Equation~(\ref{#1})} 
\newcommand{\Fig}[1]{Fig.~\ref{#1}}
\newcommand{\Figure}[1]{Figure~\ref{#1}}
\newcommand{\Sect}[1]{Sect.~\ref{#1}}
\newcommand{\Section}[1]{Section~\ref{#1}}

\newcommand{\Appendix}[1]{Appendix~\ref{#1}}

\usepackage{xspace}
\newcommand{\CII}{[C\,{\sc ii}]\xspace}

\begin{document}

\title{Spectral Data-cube Cleaning for CCAT Deep Spectroscopic Survey}

\subtitle{I. Effect of correlated noise and filtering on the power spectrum}

\author{
    A. Dev \inst{1} \orcidlink{0009-0008-6563-3681} \thanks{Corresponding Author: \email{adev@uni-bonn.de}}
	\and C. Karoumpis \inst{1} \orcidlink{0000-0003-3259-7457}
    \and Y. Okada \inst{2} \orcidlink{0000-0002-6838-6435}
    \and K. Basu \inst{1} \orcidlink{0000-0001-5276-8730}
    \and F. Bertoldi \inst{1} \orcidlink{0000-0002-1707-1775}
    \and D. Chung \inst{3} \orcidlink{0000-0003-2618-6504}
    \and J. Clarke \inst{1} \orcidlink{0009-0006-6570-5804}
    \and R. Freundt \inst{3} \orcidlink{0000-0002-8169-538X}
    \and T. Nikola \inst{3} \orcidlink{0009-0000-6722-7216}
    \and T. Oak \inst{1} \orcidlink{0009-0001-9456-1401}
    \and D. Riechers \inst{2} \orcidlink{0000-0001-9585-1462}
}

\institute{
    Argelander-Institut f\"ur Astronomie, Universit\"at Bonn,
    Auf dem H\"ugel 71, 53121 Bonn, Germany
    \and
    I.\ Physikalisches Institut, Universit\"at zu K\"oln,
    Z\"ulpicher Stra\ss e 77, D-50937 K\"oln, Germany
    \and
    Department of Astronomy, Cornell University, 
    Ithaca, NY 14853, USA
}


\abstract
{
    The Epoch of Reionization Spectrometer (EoR-Spec) on the Fred Young Submillimeter Telescope (FYST) will conduct the CCAT Deep Spectroscopic Survey (DSS). It will operate over the $210 - 420$ GHz frequency range to perform line-intensity mapping (LIM) of redshifted \CII $158 \, \mu$m fine-structure line emission at redshifts $z = 3.5 - 8.0$. 
}
{
    EoR-Spec observations are affected by atmospheric noise ($1/f$ noise) and systematics that impact LIM power spectrum (PS) measurement. We aim to quantify the impact of $1/f$ noise and data-reduction filtering on LIM PS recovery using realistic end-to-end simulations and evaluate a Filter-and-Bin (F\&B) pipeline for the EoR-Spec instrument.
}
{
    We simulated EoR-Spec observations in the $330 - 370$ GHz band ($\langle z \rangle\textsubscript{\CII} = 4.45$), including instrument response, \CII and CO emission, atmospheric noise, and observing strategy. The resulting mock timestreams were processed with the F\&B pipeline consisting of polynomial de-trending, scan-synchronous correction, common-mode subtraction, and PCA filtering, and subsequently binned into LIM data-cubes.
}
{
    The F\&B pipeline suppresses atmospheric $1/f$ noise by approximately four orders of magnitude at the lowest temporal frequencies ($<0.1\,\mathrm{Hz}$), leaving only minor residual correlated noise. In the nominal scenario of a single EoR-Spec module operating at 50\% observing efficiency, the DSS is expected to detect the combined \CII + CO PS signal on the small shot-noise dominated spatial scales ($k > 0.1\,\mathrm{Mpc}^{-1}$). Under the ideal configuration of two EoR-Spec modules operating at 100\% observing efficiency, the DSS can achieve detection at all spatial scales with the planned 2000 hours of observations. The pipeline transfer function exceeds 80\% at $k \gtrapprox 0.5\,\mathrm{Mpc}^{-1}$ but falls below 20\% at $k \lessapprox 0.1\,\mathrm{Mpc}^{-1}$, indicating substantial suppression of large-scale modes.
}
{
    The F\&B pipeline effectively mitigates $1/f$ noise and is well suited for recovering the shot-noise regime of the \CII + CO PS, while large-scale clustering measurements will require more advanced map-making methods.
}


\maketitle
%

\section{Introduction} \label{sec:introduction}
The build-up of stellar mass in galaxies is one of the key questions in observational cosmology and is traced by the cosmic star formation rate density (SFRD), which measures stellar mass assembly as a function of redshift. Although well constrained at $z \lesssim 2$ \citep{Madau2014}, the SFRD at $z>2$ remains uncertain due to incomplete galaxy samples, dust-biased tracers (e.g. UV, H$\alpha$), and observational limitations beyond cosmic noon ($z \sim 2$). Sub-millimetre (sub-mm) and far-infrared (FIR) surveys probe obscured star formation, but the faint, low-mass galaxy population remains poorly sampled \citep{Walter_Decarli_2016, Hodge_Cunha_2020, Gruppioni2020_ALPINE, Bouwens2020_ALMA, Bourne2017_SCUBA2, Algera2023_REBELS, Zavala2021_ALMA}. Flux-limited galaxy surveys are biased towards the brightest systems and suffer from cosmic variance \citep[e.g.][]{Castellano2023_GLASS_JWST, Bethermin2020_ALPINE_CII, Skelton2014_HST}. Line-intensity mapping (LIM) offers a complementary approach by measuring aggregate emission from all sources using spectral lines such as \CII, CO, and HI, enabling three-dimensional (3-D) mapping of large-scale structure (LSS) across wide volumes without resolving individual galaxies \citep[see][for reviews]{Kovetz2017_LIM, Kovetz2019_LIM, Schaan2021_LIM, Chang2026_LIM}.

\CII and CO LIM provide complementary tracers of star formation and molecular gas across cosmic time. The \CII $158\,\mu$m fine-structure line is the dominant coolant of the interstellar medium \citep{Hollenbach1999_ISM, Carilli2013_ISM} and one of the brightest lines in star-forming galaxies, contributing up to $1\%$ of the total far-infrared luminosity \citep{Stacey1991_CII, Stacey2010_CII, Malhotra2001_ISM}. Its luminosity correlates with the star formation rate (SFR) over several orders of magnitude, from local galaxies to high-redshift systems \citep{DeLooze2014_CII_SFR, HerreraCamus2015_CII_SFR, GLagache2018_CII_158}, making \CII a powerful tracer of star formation and LSS. CO rotational transitions trace the molecular interstellar medium and the gas reservoir from which stars form; as the most accessible proxy for H\textsubscript{2}, CO luminosity is linked to star formation through the Kennicutt-Schmidt relation \citep{Kennicutt1998_KS, Bolatto2013_CO_H2, Solomon2005_CO, Kennicutt2012_CO_SFR}. CO LIM has been constrained by COPSS and mmIME at $z\sim1.3$ -- 3.6 \citep{Keating2016_COPSS, Keating2020_mmIME}, while COMAP targets CO($J=1$--0) at $z\sim3$ and has reported upper limits from its second observing season \citep{Cleary2022_COMAP, Stutzer2024_COMAP_Season2, Chung2024_COMAP_Season2}. Regarding \CII LIM, no direct constraints have been placed so far, with CONCERTO, TIME, TIM, and EXCLAIM targeting emission from intermediate redshifts to the Epoch of Reionization (EoR) \citep{CONCERTOCollaboration2020, Hu2024_CONCERTO, Crites2014_TIME, Yang2025_TIME, Vieira2020_TIM, Cataldo2021_EXCLAIM}. 

A particularly promising upcoming study is the  Deep Spectroscopic Survey (DSS), targeting the redshifted \CII emission at $z \sim 3.5$ to 8 and CO rotational transitions from lower redshifts. This survey will be conducted with the newly built Fred Young Submillimeter Telescope (FYST) operated by the CCAT collaboration 
\footnote{https://www.ccatobservatory.org/} \citep{CCATCollab2022_Forecasts}.
FYST is a 6-metre diameter, wide field-of-view, high mapping speed sub-mm telescope located at Cerro Chajnantor (5600 m) in northern Chile, and is uniquely positioned for high-redshift LIM experiments and wide-field sub-mm surveys. The DSS will allocate approximately 4000 hours of observing time over a five-year campaign across two deep fields: the Extended COSMOS (E-COSMOS) field and the Extended Chandra Deep Field South (E-CDFS). Each field will cover 4 deg\textsuperscript{2}, with approximately 2000 hours of observing time dedicated to each. The survey will be carried out with the Epoch of Reionization Spectrometer (EoR-Spec), which uses a cryogenic scanning Fabry-P\'erot interferometer (FPI) as its main spectral filter. The EoR-Spec is one of the seven modules of the Prime-Cam instrument \citep{EVavagiakis2018_PrimeCamfirst}, and is optimised for high-redshift \CII intensity mapping. The Prime-Cam instrument is one of the two instruments on FYST. The EoR-Spec will cover the full frequency range from 210 to 420 GHz with a \textit{spectral resolving power} of $R = 100$. 

Astronomical observations at millimetre (mm) and sub-mm wavelengths are strongly affected by atmospheric emission, primarily from water vapour, which both attenuates astronomical signals and introduces brightness temperature fluctuations. Observations are therefore limited to atmospheric transmission windows. The high-altitude Cerro Chajnantor site, with exceptionally low precipitable water vapour (PWV), provides favourable conditions, with PWV levels $\sim 28\%$ lower than the nearby Chajnantor plateau \citep{Cortes2020_Chajnantor, Radford2016}. Spatial and temporal variations in water vapour, driven by atmospheric turbulence, introduce correlated low-frequency noise in detector timestreams. This behaviour has been characterised through total intensity studies that model atmospheric turbulence and its impact on mm and sub-mm observations \citep[e.g.][]{Church1995, Lay2000_atm_cmb, Errard2015_Atm_model, Sayers2010_atm, Morris2022_ACT, Morris2025_ACT}. As most emission originates from the lower atmosphere, it appears as a common-mode signal across detectors and is typically mitigated using common-mode subtraction, time-domain and Fourier-mode filtering. However, such processing can affect the recovery of the LIM signal and must be carefully quantified.

Several studies have forecast the detectability of the \CII LIM signal with the CCAT DSS for a range of redshift ranges \citep[e.g.][]{DChung2020_Forecasting_CII, CCATCollab2022_Forecasts, Padmanabhan2019_CII_LIM, CKaroumpis2022_CII_Lineintensity, JClarke2024_CII_luminosity, Marcuzzo2025_CII_LIM, Horlaville2024_CII_LIM} and investigated the impact of CO interloper foregrounds from lower-redshift galaxies \citep[see][]{CKaroumpis2024_CII_Lineintensity, ARoy2024_Crosscorrelation}. However, these studies have generally considered only the instrumental white-noise component. Quantifying the impact of low-frequency atmospheric noise, which can contaminate the largest spatial modes, is important for understanding the recovery of the LIM signal on large scales. In this work, we studied the impact of atmospheric noise on EoR-Spec observations by generating realistic detector timestream simulations for a range of observed frequencies. We then developed a data-reduction pipeline for the EoR-Spec dataset, applied it to the mock observations, and evaluated the effect of correlated noise and filtering on the recoverable LIM signal.

This paper is organised as follows: \Sect{sec:end-to-end-simulation-framework} describes the modelling of the EoR-Spec instrument and the simulation framework developed in this work to generate the EoR-Spec mock data. In \Sect{sec:data-reduction-pipeline}, we introduce the different steps of the data-reduction pipeline and apply it to the simulated dataset. We then make spectral maps from the cleaned dataset and generate the LIM data-cube. \Section{sec:results} presents the results of the filtering pipeline on the correlated noise characteristics and the sensitivity forecasts for the DSS taking into account the correlated noise. We also analyse the effect of the filtering on the recoverable LIM signal through the transfer function. Finally, we summarise the work and conclude in \Sect{sec:conclusion}.


\section{End-to-end Simulation Framework}\label{sec:end-to-end-simulation-framework}
In this section, we describe the end-to-end simulation framework used in this work. The framework accounts for the instrument focal plane, observation schedule, telescope scan pattern, input astrophysical signal, and atmospheric loading. EoR-Spec instrument modelling is presented in \Sect{sec:eorspec-instrument-modelling} and \Sect{sec:eorspec-spectral-binning}, followed by the methodology for generating mock observations in \Sect{sec:creating-mock-observations}. A flowchart demonstrating the simulation framework and data flow is included in \Fig{fig:eorspec_data_flowchart}.

\subsection{EoR-Spec Instrument Module}\label{sec:eorspec-instrument-modelling}
\begin{figure*}
\centering
  \includegraphics[width=17cm]{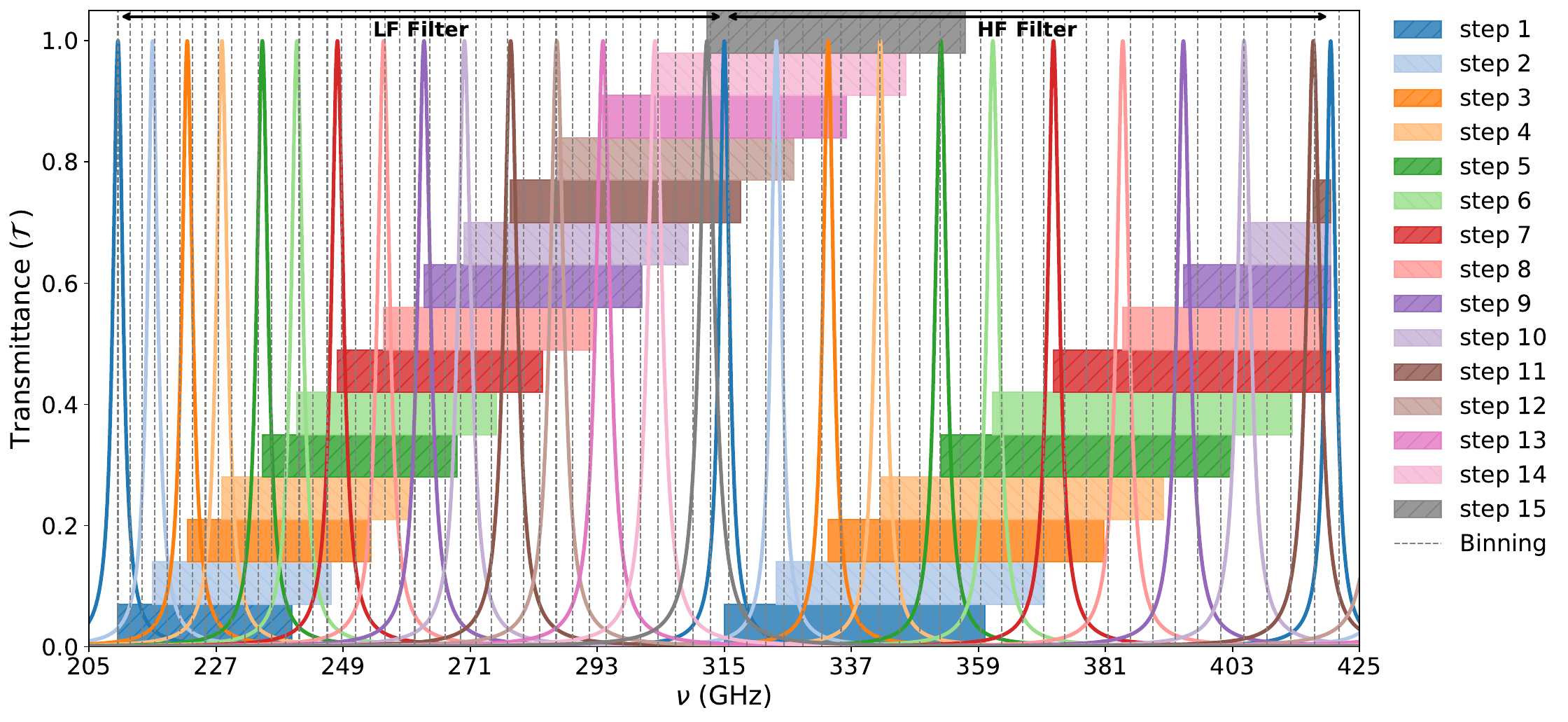}
  \caption{Transmittance ($\mathcal{T}$) showing the ideal spectral transmission profile, without accounting for losses and absorptance, for the 15 discrete FPI steps of EoR-Spec, is shown for normal incidence ($\theta = 0^{\circ}$). Shaded regions indicate the spectral coverage at each FPI step due to the range of incident rays. Vertical dashed lines show the spectral binning for EoR-Spec (see \Sect{sec:eorspec-spectral-binning}). The low-frequency (LF) and high-frequency (HF) filter bands are indicated, covering the full spectral window of EoR-Spec.}
  \label{fig:fpi_spectral_full}
\end{figure*}

The EoR-Spec module features three focal plane arrays comprised of Microwave Kinetic Inductance Detectors (MKIDs). Two of these arrays, known as the Low-Frequency Array (LFA), are centered at 260 GHz and cover the 210–315 GHz spectral range. The High-Frequency Array (HFA), which covers the higher frequencies from 315-420 GHz is centered at 365 GHz \citep{CCATCollab2022_Forecasts}. Each of the two LFA arrays contains 1728 MKIDs, while the single HFA array is populated with 3072 MKIDs. This configuration enables simultaneous spectral multiplexing across multiple frequency bands while maintaining approximately uniform sensitivity across the observed spectral range.

The FPI, also referred to as an etalon, is mounted at the Lyot stop of the EoR-Spec module. The Lyot stop is located at the pupil plane of the EoR-Spec optical system, suppressing stray light and defining the beam illumination \citep[see][for descriptions of the instrument optical design]{RFreundt2024_CCATstatusupdate,NCothard2020_DesignCCATprime, ZHuber2022_CCATprimeoptical}. The FPI transmits light onto the three focal-plane arrays of the EoR-Spec instrument, covering the $1.3^{\circ}$ field of view of the camera module. Due to conservation of optical throughput, the angular offset on the sky corresponding to a detector position is related to the beam incidence angle at the Lyot stop by:
\begin{equation}
    \label{eq:lyot_fpi}
    \theta_\mathrm{sky} \, D_\mathrm{tel} = \theta \, D_\mathrm{Lyot}
\end{equation}
where $\theta_\mathrm{sky}$ is the angular separation on the sky from the centre of the EoR-Spec module, $D_\mathrm{tel}$ is the telescope diameter ($\approx 5.8\,\mathrm{m}$), $\theta$ is the beam incidence angle at the FPI, and $D_\mathrm{Lyot}$ is the diameter of the Lyot stop ($13\,\mathrm{cm}$) where the FPI is installed. 

An FPI consists of two identical and parallel reflective mirrors arranged to form a resonant cavity. The wavelength transmitted through an FPI is determined by the distance between its mirrors, with different spacings selecting different wavelengths for observation. The transmission profile of the etalon is given by interference between multiple internally reflected beams within the reflective mirrors. For an ideal and lossless Fabry-P\'erot system, the transmission profile can be described by:
\begin{equation}
	\label{eq:transmission2_fpi}
    \mathcal{T}(\nu \,; d,\theta) = \frac{1}{1 + F \sin^2 \left(\frac{\pi \,\Delta \nu}{{\Delta \nu}_{\mathrm{FSR}}} \right)} 
    = \frac{1}{1 + F \sin^2 \left( \, \frac{\pi \, \Delta \nu \, 2 nd \, cos \theta}{c} \right) }
\end{equation}
Here, $\mathcal{T}$ is the transmittance given as the ratio of transmitted intensity $(I_t)$ to the incident intensity $(I_i)$. \Equation{eq:transmission2_fpi} describes the ideal, lossless spectral transmission profile of a Fabry-P\'erot etalon for radiation of frequency $\nu$, incident at an angle $\theta$, given a fixed cavity gap thickness $d$ between the parallel reflective mirrors of the etalon, and $n$ is the refractive index within the cavity. The term $\Delta \nu_{\mathrm{FSR}} = \frac{c}{2nd \, cos \theta}$ is known as the \textit{free spectral range} parameter, which gives the frequency spacing between two adjacent resonance peaks of an FPI. We note that, $\theta = 0^{\circ}$ corresponds to the optical axis at the centre of the module, while $\theta_\mathrm{max}$ corresponds to the outer edge of the focal-plane arrays. This angular dependence produces a radial blue-shift in the transmitted frequency across the arrays, enabling spectral multiplexing for each FPI configuration.

We refer the reader to \Appendix{sec:fpi-theory}, \citet{born1999principles} and \citet{hecht_optics_2017} for a thorough treatment of FPI and multi-beam interference theory. As shown in \Appendix{sec:fpi-theory}, \Eq{eq:transmission2_fpi} can be expressed in the form of a Lorentzian function and therefore, the FPI has a Lorentzian spectral profile with the full width at half-maximum (FWHM) given by $\Delta \nu_\mathrm{FWHM} = \frac{\nu}{R}$.

The EoR-Spec instrument achieves spectral scanning from $210 - 420$ GHz by varying the cavity gap thickness $d$ in the FPI across 15 discrete steps, hereafter referred to as the 15 FPI steps (additional details provided in \Appendix{sec:detector-counts}). The FPI operates simultaneously in the second and third orders of interference, allowing multiple frequency bands to be observed at each FPI step. The ideal spectral transmission profile for EoR-Spec, without accounting for absorptance and losses, is shown in \Fig{fig:fpi_spectral_full} for all 15 steps. The profile at each FPI step is a function of incidence angle $\theta$ (shown by the shaded regions), as described by \Eq{eq:transmission2_fpi}. 

\subsection{EoR-Spec Spectral Binning}\label{sec:eorspec-spectral-binning}
\begin{figure}
  \resizebox{\hsize}{!}{\includegraphics{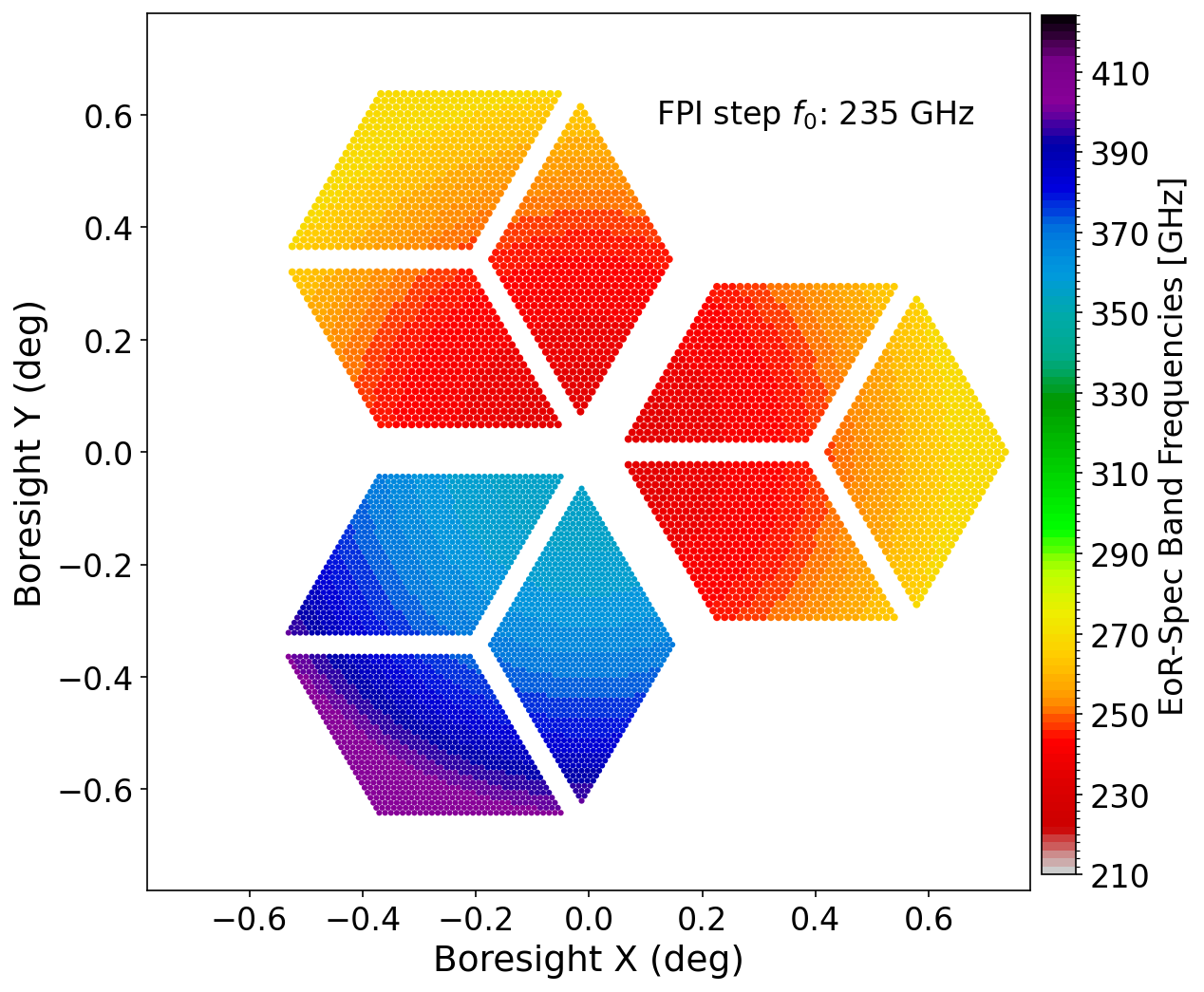}}
  \caption{EoR-Spec focal plane array layout projected onto the boresight coordinate plane. The cavity gap  corresponds to the $5\textsuperscript{th}$ FPI step, tuned to a resonance frequency of 235 GHz, associated with the second-order resonance at normal incidence. The layout shows two low-frequency arrays (LFA: red -- yellow), operating at the second-order resonance, and a single high-frequency array (HFA: blue -- purple), targeting the third-order FPI resonance \citep{CCATCollab2022_Forecasts, RFreundt2024_CCATstatusupdate}. The colour scale shows the full EoR-Spec spectral coverage, with groups of detectors sampling different spectral bins, based on a discretely binned frequency grid. The radial variation in observed frequency across each array is due to the incidence angle dependent nature of the FPI transmission, resulting in a blue-shift as the angle increases radially outward.}
  \label{fig:fp_eorspec_step235}
\end{figure}

The EoR-Spec instrument observes a continuous distribution of frequencies associated with the low and high frequency bands across the three arrays. We binned the total frequency coverage into discrete spectral bins defined by the peak resonance frequencies of the FPI. For each detector and FPI step, the corresponding peak transmission frequency was identified and assigned to the spectral bins. The bin width was chosen to be equal to the $\Delta \nu_{\mathrm{FWHM}}$ of the FPI spectral profile. In this approach, a measurement corresponding to a specific FPI step and detector position is associated with a single spectral bin, while ignoring contributions from neighbouring bins. An example of this binning method for the three arrays is shown in \Fig{fig:fp_eorspec_step235} for the spectral coverage of the $5^{\mathrm{th}}$ FPI step. 

The CCAT DSS is designed to observe the E-COSMOS and E-CDFS fields for 2000 hours each, with the EoR-Spec instrument over the nominal five-year survey period \citep{CCATCollab2022_Forecasts}, covering the $210 - 420$ GHz spectral range by stepping through 15 discrete FPI steps. To achieve uniform depth per spectral bin, the baseline plan is to observe the fields with equal integration time allocated to each FPI step. As a result, the integration time per FPI step for a given field is $\frac{2000}{15} \,\mathrm{hours} \approxeq 133 \,\mathrm{hours}$. 

For a given spectral bin, let \(d_i\) denote the number of contributing detectors from the \(i^{\mathrm{th}}\) FPI step observed for a duration \(t_i\). The total integration-time-weighted contribution to the bin is then proportional to \(\sum_{i=1}^{n} t_i d_i\). For EoR-Spec, \(n=15\) and each FPI step is observed for approximately equal duration (\(t \approx 133\) h), reducing this expression to \(t \sum_{i=1}^{n} d_i\). As shown in \Fig{fig:dets_per_FPIchnl}, the cumulative contribution from all FPI steps corresponds to roughly \(1000\) detectors per spectral bin for a single EoR-Spec module.

\subsection{Creating Mock Observations}\label{sec:creating-mock-observations}
We used a simulation framework based on Time-Order\-ed Astro\-physics Scalable Tools (TOAST)
\footnote{TOAST \url{https://hpc4cmb.github.io/toast/} \\
TOASTv3.0.0a29 was used for this work \\ \url{https://github.com/hpc4cmb/toast/releases/tag/3.0.0a29}}
\citep{Kisner2023_TOAST} within this work to generate the mock observations for the EoR-Spec instrument. TOAST is a software package used to simulate and process data from mm and sub-mm telescopes that collect detector time\-streams, also referred to as Time-Ordered Data (TOD). 
TOAST supports scalable workflows optimized for large datasets, enabling simulation of detector responses, atmospheric and instrumental systematics, and the re\-construction of sky maps using various filtering and map-making methods. The TOAST software framework has been used in several studies for time-domain simulation, as well as map-making in ground  and space-based mm and sub-mm experiments \citep[e.g.][]{Puglisi2021_TOAST, Dachlythra2024_SO_TOAST, Natoli2018_CORE_spaceTOAST, Ma2023_TOLTEC_TOAST}. 

In this study, we included a range of observational and instrumental parameters into our simulations for the EoR-Spec instrument, including the spectral bins of the instrument, simulating multiple FPI steps, the focal plane layout, effect of the beam, telescope on-sky scanning strategy, observing schedule, targeted sky region, and site-specific details like location, altitude, and weather conditions. As described in \Sect{sec:simulating-atm-realisation}, the end-to-end simulations produced mock datasets with correlated atmospheric noise, allowing us to evaluate the data-reduction pipeline and its impact across different spatial scales. This detailed TOD (or detector timestream) simulation of the DSS field for FYST, using the mock EoR-Spec configuration, was based on generating mock detector timestreams from astrophysical forecast input maps and incorporating atmospheric and instrumental noise models. This approach, unlike map-based forecasts \citep[e.g.][]{CKaroumpis2022_CII_Lineintensity, JClarke2024_CII_luminosity,DChung2020_Forecasting_CII}, was built on simulating realistic TOD for each detector and allowed us to investigate the impact of correlated noise, in addition to the white noise component, on the LIM science case.

To reduce the computational cost of the simulations, we considered 5 of the 15 FPI steps, covering the 330--370 GHz spectral range within the HFA band. The HFA range was chosen because atmospheric noise contamination is more challenging at higher observing frequencies than in the LFA band. This subset was selected to maintain an approximately uniform cumulative detector count of approximately 500 detectors per spectral bin, corresponding to roughly 50\% of the full 5-year DSS dataset from a single EoR-Spec module over the chosen spectral range. The reduced dataset was sufficient to demonstrate the analysis methodology and evaluate the impact of the data-reduction pipeline on the recoverable power spectrum. To simulate observations, we implemented a mock observing schedule for the E-COSMOS field using the selected FPI steps. The schedule spanned the years 2027 -- 2031, consistent with the planned timeline of the CCAT DSS. The FPI was stepped every 5 minutes while cycling through the five selected FPI settings, with each step accumulating approximately 133 hours of integration time.

\subsubsection{Astrophysical Signal Forecasts and Input Signal Maps}\label{sec:input-signal-maps}
In the context of the CCAT DSS, \CII and CO line transitions are the primary signal targets for the LIM science case \citep{CCATCollab2022_Forecasts}. The \CII emission from galaxies at redshifts $z \sim 4.1 - 4.8$ is redshifted into the sub-mm window considered in this study, while CO emission from lower-redshift galaxies ($z \lesssim 3.2$) enters the same band as a foreground \citep[see Table 1 in][hereafter K24]{CKaroumpis2024_CII_Lineintensity}. Although CO acts as a foreground for the high-redshift \CII signal, the tomographic structure of CO is also an important LIM science target, providing constraints on the cosmic molecular gas content at intermediate redshifts \citep{Chung2022_COMAP_Season1, Chung2024_COMAP_Season2}. In this work, we included tomographic line-intensity maps from both \CII and CO lines within the $330 - 370$\,GHz spectral range as input astrophysical LIM signal for the end-to-end simulation. We did not distinguish between the targeted \CII astrophysical LIM signal and the CO astrophysical foregrounds, treating them collectively as a single astrophysical signal. We then analysed the impact of the atmospheric foregrounds and instrumental noise systematics on this combined signal.

The input \CII tomographic mock maps were based on the fiducial predictions of \citet[hereafter K22]{CKaroumpis2022_CII_Lineintensity}. They began by constructing a dark matter light-cone by combining the Illustris TNG300-1 halo catalogues \citep{Pillepich2018_TNG300, Donnari2019_TNG300} from several simulation snapshots, extending from $z = 0$ to $z = 10$. In the fiducial model of K22, the intrinsic SFR of the TNG300 simulated galaxies were recalibrated using TNG100 to account for the low mass resolution of the 300 Mpc box simulation. Then the SFR for each galaxy was translated to \CII line luminosity ($L\textsubscript{\CII}$) using a SFR-L\textsubscript{\CII} relation. For this work, we used the fiducial \CII tomographic data-cube product corresponding to the SFR-L\textsubscript{\CII} relation from \citet[hereafter V15]{LVallini2015_CII_SFR_rel}. Among the models presented in K22, the fiducial recalibrated TNG300 and V15-based mock maps provide intermediate \CII brightness between the optimistic and pessimistic extreme scenarios, differing by about one order of magnitude in mean intensity across a given redshift (see Fig.~6 in K22).

The mock CO tomographic mock maps used in this work were developed by K24. These maps were generated from the above described TNG300-based light-cone, where each galaxy was assigned a CO(1--0) luminosity using SFR-L\textsubscript{CO} scaling relations. Higher rotational transitions, up to $J\textsubscript{up}=12$, were inferred by applying a spectral line energy distribution model. The mock CO intensity maps were generated by summing the CO line emission from transitions $J\textsubscript{up}=3$ to 12, providing the total CO emission contribution within each spectral bin. We combined the CO and \CII intensity mock maps for each spectral bin in the $330 - 370$ GHz spectral range to produce the total line-intensity input mock maps for this work. The \CII and the total CO emission within the $330 -370$ GHz spectral range have comparable mean intensities, with both K24 and \citet{Bethermin2022_CONCERTO} predicting a cross-over frequency of $\sim 360$ GHz, above which the \CII signal becomes the dominant contributor among the two signals (see Fig. 7 in K24).

As discussed in \Sect{sec:eorspec-instrument-modelling}, the FPI has a Lorentzian spectral profile. The combined \CII and CO input mock maps were defined on a spectral grid with a spacing of one-third of the Lorentzian FWHM, satisfying the Nyquist sampling criterion. We performed spectral convolution of the input signal model along the line-of-sight to match the instrument spectral profile. For the input mock maps covering the $330 - 370$ GHz range, we used $\Delta \nu_\mathrm{FWHM} = 3.5 \,$ GHz of the spectral profile as the kernel for the convolution operation. The spectral smoothing has the effect of spreading the signal power across neighbouring spectral bins. The spatial pixel scale on the combined input mock maps were set to one-third of the instrument beam FWHM to meet the Nyquist sampling criterion. We accounted for spatial smoothing due to the instrument beam, by convolving the spectrally smoothed mock maps with a Gaussian kernel representing the beam, with the beam FWHM at 350\,GHz for EoR-Spec being $37\,\arcsec$ \citep{CCATCollab2022_Forecasts}. 

\begin{figure}
  \resizebox{\hsize}{!}{\includegraphics{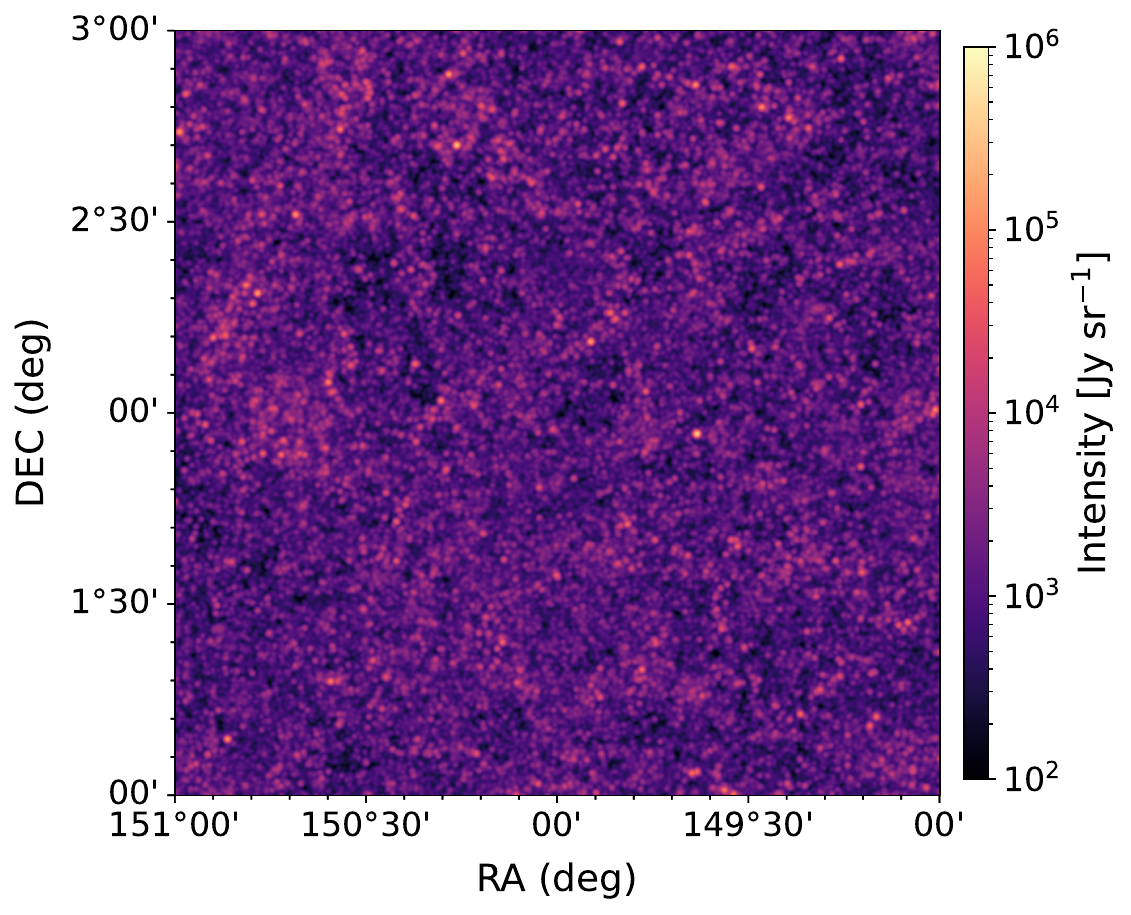}}
  \caption{A $2^{\circ} \times 2^{\circ}$ cutout of the combined \CII and CO line-intensity mock map at the 365\,GHz spectral bin ($z\textsubscript{\CII} \sim 4.2$), based on K22 and K24. The map is convolved with the instrument beam and spectral response, and projected onto the E-COSMOS field using the CAR projection.}
  \label{fig:input_CO_CII_2deg}
\end{figure}

We used a plate carr\'ee (CAR) projection for the tomographic maps, centered on the E-COSMOS field (RA = 150.0\,$^\circ$, Dec = 2.0\,$^\circ$). An example of a spectrally and spatially smoothed $2^{\circ} \times 2^{\circ}$ cutout of the combined input mock map at 365\,GHz, centred on the E-COSMOS field, is shown in \Fig{fig:input_CO_CII_2deg}. In practice, we used a larger $4^{\circ} \times 4^{\circ}$ input signal map for scanning to ensure complete coverage of map pixels near the field boundaries (see \Fig{fig:coadd_hits}). Only the central $2^{\circ} \times 2^{\circ}$ region was used for power spectrum analysis. Such combined input mock maps were produced for all spectral bins within $330 - 370$ GHz, and 11 of these maps were used to construct the data-cube for scanning with the simulated detectors and mock observing strategy (described in \Appendix{sec:scan-pattern}).

\subsubsection{Simulating Atmospheric Realisation}\label{sec:simulating-atm-realisation}
The TOAST framework simulates atmospheric emission following a 3-D turbulence model of water vapour fluctuations that captures both the spatial and temporal correlations in the signal as seen by different detectors. The atmosphere is simulated as a slab of turbulent cells based on Kolmogorov model of turbulence with specified injection and dissipation scales that define the range of eddy sizes contributing to brightness fluctuations \citep{Kolmogorov1941, Tatarski1961}. The slab is moved across the field of view of the focal plane by a constant horizontal wind speed, producing realistic time-variable correlated patterns in the TOD that depend on detector angular separation, scan speed and wind direction. The resulting fluctuations produce a large-scale, low-frequency noise component, called as the $1/f$ noise. The atmospheric simulation uses a 3-D and turbulence-driven model introduced in \citet{Errard2015_Atm_model}, which describes the mm and sub-mm brightness fluctuations arising from water-vapour inhomogeneities in the atmosphere using a physical model. Each detector measures the cumulative emission along its line of sight, integrating the brightness fluctuations through the turbulent layers of the atmosphere. A weather file for the Atacama-site built from the MERRA-2 historical dataset is used to draw the PWV and weather parameters data required for the simulation \citep{Gelaro2017_MERRA2}. The \texttt{GenerateAtmosphere} operator is used from the TOAST framework to generate the realisations and atmospheric volumes , and \texttt{ObserveAtmosphere, SimAtmosphere} operators are used to observe through a simulated atmospheric volume and generate the atmospheric  TOD component ($n_i^{\mathrm{atm}}(t)$) for detectors respectively. For this work, two atmospheric realisations were simulated. A coarse component was used to capture the large-scale turbulence that are usually responsible for producing the slowly varying drift patterns seen in sub-mm data, and a fine component to model the small-scale fluctuations and short timescale variations. Both are integrated along detector line of sight to produce realistic, multi-scale atmospheric emission in the simulated TOD. The two components are defined by different spatial turbulence scales and as a result, contribute to different temporal scales in the TOD. The coarse component primarily captures large-scale atmospheric structures that typically produce fluctuations on timescales ranging from several minutes to hours, while the fine component models smaller-scale atmospheric variations that contribute fluctuations on timescales of seconds to minutes. The parameters used for this atmospheric simulation are given in Table~\ref{tab:atm_params}. 

\subsubsection{Simulating Detector Timestream}\label{sec:simulating-detector-timestream}
We used the TOAST framework to convert the input (\CII and CO) mock maps into signal-only detector timesteams, following the predefined observation schedule and scan pattern. This was done using the pixelization (\texttt{PixelsWCS}) and scanning (\texttt{ScanWCSMap}) operators. The telescope boresight and detector pointing were computed by projecting the Az-El scan pattern onto the sky in the RA-DEC coordinates. We adopted a Stokes-I-only model throughout, as EoR-Spec is a total-power imaging spectrometer. The scanning operator sampled the mock map at each instantaneous sky position and recorded the corresponding intensity as the detector signal, while flagging samples falling outside the field footprint. This pipeline produced signal-only TOD, to which we subsequently added atmospheric and instrumental noise.
The white-noise component of the simulated TOD was based on the detector sensitivity values reported for the DSS in \citet{CCATCollab2022_Forecasts} (see Table~1c).

Each sample $t$ corresponds to a single detector measurement recorded at the sampling rate $f_s$. The total number of samples in a timestream is $N_\mathrm{samples} = f_s \, T_\mathrm{obs}$, where $T_\mathrm{obs}$ is the observation time of one observation scan. The TOD  model for the $i\,\textsuperscript{th}$ detector is given by:
\begin{equation}
	d_i(t) = \sum_p P_{ip}(t)\, m_p + n_i^{\mathrm{atm}}(t) + n_i(t) \\
\end{equation}
where $m_p$ is the input map intensity in map pixel $p$, $P_{ip}(t)$ is the pointing matrix element that selects the observed map pixel $p$ at time $t$, and $n_i^{\mathrm{atm}}(t)$ and $n_i(t)$ denote the atmospheric and instrumental noise components, respectively. This mock TOD is used to test the pipeline and understand the effect of $1/f$ noise in the context of LIM-experiments from sub-mm telescopes. 


\section{Data Reduction Pipeline}\label{sec:data-reduction-pipeline}
\begin{figure*}
\centering
  \includegraphics[width=17cm]{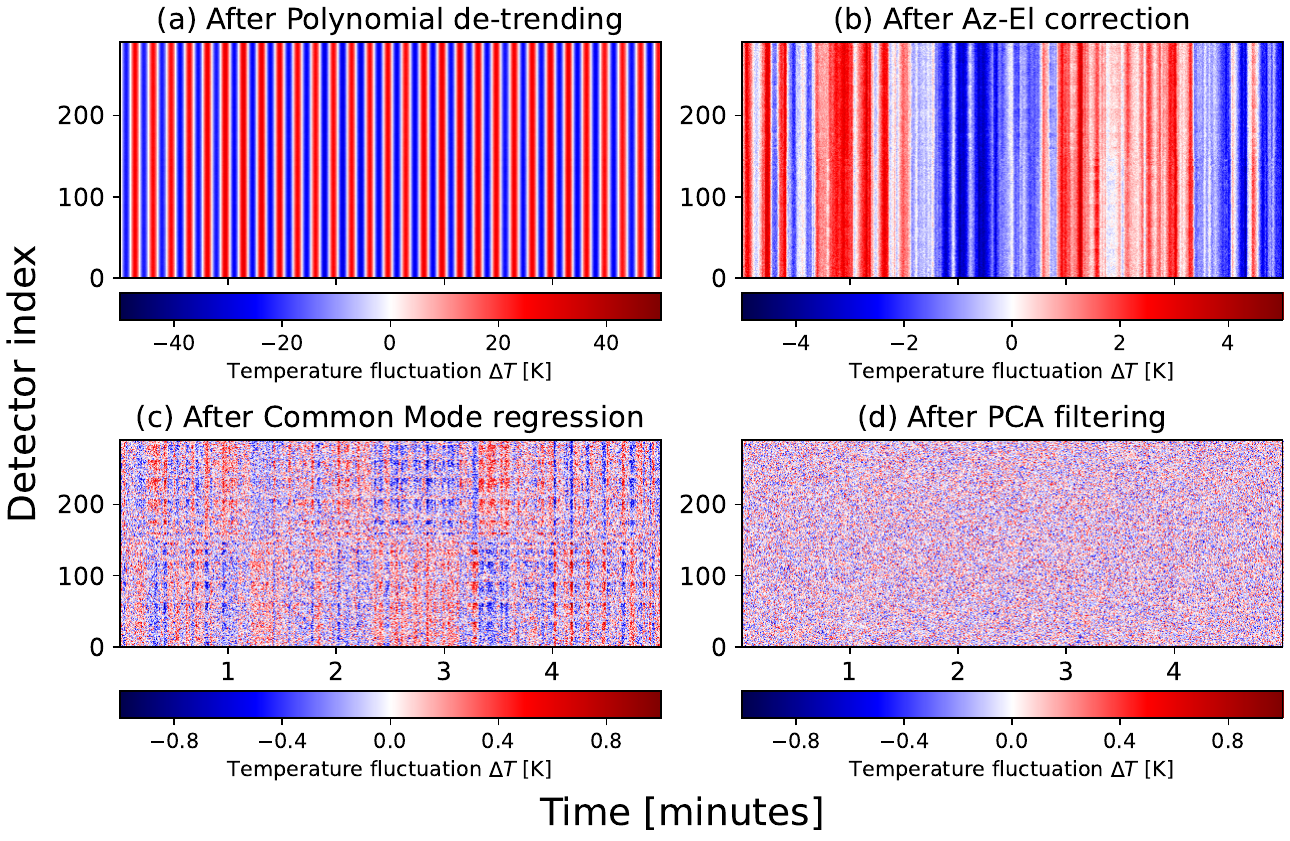}
  \caption{Visualisation of TOD for simulated detectors for a single FPI step during $\sim 5$ minutes of observation after successive stages of data cleaning. Panel (a) shows the TOD after polynomial de-trending, which removes the global drifts and leaves variations dominated by the telescope motion. Panel (b) displays the data after correcting for the telescope Az-El motion, after which the TOD is dominated by common mode fluctuations. Panel (c) shows the data after common mode regression has been applied and most of the correlated $1/f$ modes across the detectors have been regressed out. The residual $1/f$ fluctuation that could not be modelled by the common mode filter is seen at this stage. Panel (d) presents the TOD after removing the leading 3 principal components, which suppresses the residual correlated structure. The data, after applying all filtering stages, is dominated mostly by white noise as seen in Panel (d). For each panel, the data was normalised by the maximum absolute value of each dataset to represent all stages of the pipeline on the same colour scale.}
  \label{fig:2d_tod_clean}
\end{figure*}

The fundamental data product in mm and sub-mm instruments using KID arrays is the detector timestream. Each detector samples the sky along the telescope scan pattern and records a signal proportional to the incident optical power as a function of time. These timestreams are subsequently calibrated and processed through a data-reduction pipeline to mitigate instrumental and atmospheric systematics before map-making. The choice of pipeline architecture and map-making method depends on the science goals and varies across experiments.

In this section, we describe the \emph{Filter-and-Bin} (hereafter F\&B) pipeline used to produce cleaned maps from the mock timestreams. We assumed the simulated timestreams were already well-calibrated and converted to units proportional to sky brightness temperature, and applied a sequence of filters to mitigate atmospheric and instrumental systematics before map-making.
The mock TOD were stored in observation files having samples equivalent to $ \sim 5$ minutes of observation time each, because we changed the FPI step every 5 minutes. We processed the TOD from each FPI step independently, generating maps for each spectral bin at every FPI step. This approach avoids potential mixing of timestream systematic effects associated with different FPI steps. Since each spectral bin is observed at multiple FPI steps, the corresponding independently processed maps were co-added to produce the final map in that spectral bin.

\subsection{Polynomial de-trending}\label{sec:polynomial-filtering}
In this mock dataset, since each observation unit is about 5 minutes in duration, the data is mostly dominated by telescope motion related systematics and atmospheric $1/f$ noise and we assumed a noise dominated regime. We approached the timestream by performing targeted filtering that allowed us to build a filter chain. The first step in this filtering chain was to remove the slowly varying, very low temporal-frequency ($<0.1$ Hz) drift pattern caused primarily by atmospheric emission and instrumental drifts. This global drift pattern can be modelled as a low-order polynomial. To suppress large-scale temporal drifts in each detector, we applied a polynomial de-trending per detector timestream. Each detector timestream was first down-sampled by averaging every 1000 consecutive samples to suppress high-frequency fluctuations. In order to model and remove the large-scale temporal baseline, we fitted a third-order polynomial to each down-sampled detector timestream. The polynomial represents the slowly varying component of the signal caused by atmospheric and instrumental drifts and is written as:
\begin{equation}
	p_i(t) = \sum_{k=0}^{3} a_{k,i}\ t^k \\
\end{equation}
where the coefficients were determined through a least-squares fit to the down-sampled data. The fitted trend was then evaluated at the full sampling rate and subtracted from the original timestream, thereby removing the low-frequency drift while preserving the scan-synchronous, common-mode and higher-frequency components. 

After subtraction, each detector timestream fluctuated around zero, indicating that the mean and other very low-frequency components were effectively removed. Panel (a) in \Fig{fig:2d_tod_clean} shows the de-trended TOD for an example $\sim 5$-minute observation segment, which exhibits alternating patterns associated with the temperature fluctuations induced by the Az-El telescope motion. The same behaviour is shown in panel (a) of \Fig{fig:1d_tod_clean}, for a subset of a few randomly selected detectors within the dataset, demonstrating that the timestreams are flattened and have comparable amplitudes across the dataset. This de-trending step ensured zero-centred TOD fluctuations, improving the stability and reliability of the subsequent fitting steps across diverse observing conditions.

\subsection{Correcting for telescope motion}\label{sec:correcting-for-telescope-motion}
The next dominant systematic is linked to the telescope motion, turnarounds and scan pattern. The elevation-modulated azimuth scan pattern (described in \Appendix{sec:scan-pattern}) chosen for generating this mock observational dataset varies both in elevation and azimuth. This introduced a scan-synchronous systematic contribution, which varies as a function of the boresight elevation, azimuth and time. A similar systematic signal was also seen in the Lissajous scans of the COMAP Early Science dataset \citep[see][hereafter F22]{MFoss2022_COMAPdata}. The scan-synchronous signal is dependent on atmospheric loading, observing conditions and elevation-dependent variations from the scan pattern. It produces low-frequency structures that must be removed prior to applying the subsequent steps in the filter chain.

We removed the elevation/azimuth-synchronous pickup by fitting and subtracting a sky-loading and azimuth model from the TOD of each detector. We used an empirical scan-synchronous model ($m_i^{\ \text{az-el}}(t)$), similar to F22, to fit the detector timestreams:
\begin{equation}
	m_i^{\ \text{az-el}}(t) = \frac{A_i}{\sin \left(el(t)\right)} + B_i\ az(t) + C_i
\end{equation}
where $el(t)$ and $az(t)$ are the elevation and azimuth of the boresight at time $t$, and $A_i$, $B_i$ and $C_i$ are fitting constants. These parameters were estimated by least-squares regression and the estimated model was then subtracted from each detector timestream. The Az-El corrected timesteam data $d_i^{\ \text{az-el-corr}}(t)$ is:
\begin{equation}
	d_i^{\ \text{az-el-corr}}(t) = d_i(t) - m_i^{\ \text{az-el}}(t)
\end{equation}
which suppresses low-frequency scan-synchronous structure in the data while retaining sky-synchronous fluctuations. Panel (b) of \Fig{fig:2d_tod_clean} shows the mock dataset after applying the Az-El motion correction. This step removed the alternating telescope motion related pattern seen in Panel (a), leaving the correlated noise as the remaining dominant systematic. The effect of the scan-synchronous signal correction on a subset of 10 detectors is shown in panel (b) of \Fig{fig:1d_tod_clean}, where the residual fluctuations show strong correlations across all detectors. This step also reduced the fluctuation amplitude by nearly an order of magnitude compared to panel (a).

\subsection{Common mode regression}\label{sec:common-mode-regression}
Once the scan-synchronous signal is removed, correlated structures shared across detectors become apparent. This common-mode signal arises primarily from atmospheric emission fluctuations and is a dominant contributor to the low-frequency $1/f$ noise. The common-mode (CM) filter acts as a spatial filter, suppressing large-scale correlated modes across the detector field of view while preserving smaller-scale astronomical signals.

We used a CM-filtering algorithm based on the \texttt{CommonModeFilter} operator in TOAST to remove correlated structures shared across detectors. For each FPI step and spectral bin, a CM template was constructed by averaging the detector timestreams contributing to that spectral bin:
\begin{equation}
	C(t) = \frac{1}{N_\text{det}} \sum_{i=0}^{N_\text{det}} d_i^{\ \text{az-el-corr}}(t) .
\end{equation}
For each detector \(i\), we then performed a linear regression of the timestream against the CM template, fitting a model of the form
$m_i^{\ \text{CM}}(t) = \alpha_i + \beta_i C(t)$,
where  $\alpha_i$ and  $\beta_i$ are the fitted regression coefficients. The fitted model was then subtracted from each detector timestream:
\begin{equation}
	d_i^{\ \text{CM-corr}}(t) = d_i^{\ \text{az-el-corr}}(t) - m_i^{\ \text{CM}}(t).
\end{equation}

\Figure{fig:2d_tod_clean}(c) shows the dataset after being processed by the CM-filter, which removed most of the dominant common structures present in panel (b). The corresponding  CM-filtered timestreams for a subset of detectors are shown in \Fig{fig:1d_tod_clean}(c), where the fluctuation amplitude is reduced by an additional factor of $\sim 5$ and the dataset was significantly whitened. We found that after applying the CM-filter, although most of the correlated structures were removed from the timestreams, the data still showed some residual $1/f$ noise. These underlying fluctuations, which are almost an order of magnitude smaller than the CM variations could not be captured by the CM-filter indicating the presence of additional unmodelled low-frequency noise. 

\subsection{PCA filter}\label{sec:pca-filter}
We used a filter based on principal component analysis (PCA) to pick up residual correlated structures across detectors that could not be modelled as common mode. PCA technique identifies orthogonal directions (principal components) in the detector-data space that capture the largest fraction of the total variance. It is typically used to reduce the dimensionality of datasets containing many interrelated variables, while preserving as much of the original variance as possible \citep{pearson1901,hotelling1933,jolliffe2011principal}. PCA is an unsupervised learning technique that relies solely on the statistical structure of the input data and its optimisation criterion is the maximisation of variance. As it operates directly on the data, PCA is particularly useful in our problem of identifying residual correlated systematics that are difficult to model with simple linear approaches.

In most applications where PCA is used for dimensionality reduction, the leading components that explain the largest fraction of the variance are retained, while the remaining components dominated by noise are discarded. However, in our case, the leading components are primarily composed of correlated structures that do not contain the astrophysical signal and we used PCA for data-cleaning rather than for dimensionality reduction. We therefore adopted the strategy of removing the first few dominant components while retaining the rest to preserve the astrophysical signal as much as possible. Similar approaches for using PCA in data-reduction for mm/sub-mm astronomical surveys were demonstrated by F22, \citet{EChapin2013_SCUBA2iterative} and \citet{JAguirre2010_Bolocam}.

The detector timestreams were arranged into a matrix \(D \in \mathbb{R}^{N_{\mathrm{det}} \times N_t}\), where rows correspond to detectors and columns to time samples. PCA was then applied by computing the detector covariance matrix (\(C = DD^\top\)) and identifying its dominant eigenmodes, which capture the largest correlated variance across detectors and are primarily associated with residual atmospheric fluctuations. A selected number of leading principal components (\(n_{\mathrm{comp}}\)) were reconstructed in the detector-time domain as:
\begin{equation}
D_{\mathrm{removed}} = V_L T_L^\top 
\end{equation}
where \(V_L\) contains the dominant detector eigenmodes and \(T_L = D^\top V_L\) their corresponding time-dependent amplitudes. These reconstructed modes were subtracted from the timestreams to obtain the PCA-filtered data:
\begin{equation}
D_{\mathrm{PCA-filtered}} = D - D_{\mathrm{removed}} 
\end{equation}

The choice of the number of the leading $n_{\mathrm{comp}}$ to remove depends on the level of residual systematics remaining after the PCA-filtering step. For the mock dataset simulated in this work, we found that removing 3 to 4 components was sufficient to remove the correlated $1/f$ noise. The goal was to remove the minimum number of components to approach an acceptable level of flat noise spectrum, since subsequent filtering increases the risk of astrophysical signal loss. We explored three separate cases of PCA filtering, removing $n_{\mathrm{comp}}=3$ (nominal case), $n_{\mathrm{comp}}=2$ and $n_{\mathrm{comp}}=4$, respectively on the same mock dataset. We compare the results of these filtering cases in \Sect{sec:transfer-function-analysis}. \Figure{fig:2d_tod_clean}(d) and \Fig{fig:1d_tod_clean}(d) show the dataset and the corresponding timestreams for a subset of 10 detectors after PCA filtering respectively, for the fiducial case in which the first three principal components were removed. The residual correlated components remaining after CM-filtering were captured by the leading principal components, and the PCA-filtered dataset is mostly dominated by white noise. The aim of the filtering steps was to remove $1/f$ correlated noise while preserving the underlying signal, and therefore we avoided aggressive filtering at each stage.

\subsection{Map-Making}\label{sec:map-making}
Map-making is the process of reconstructing a pixelised representation of the observed signal from the calibrated detector timestreams. In the previous sections, we described the methods to simulate the calibrated dataset for the EoR-Spec instrument and the individual filters used in the data-reduction pipeline to clean the systematics and correlated noise in the mock dataset. The final remaining step before performing scientific analysis on the data is to produce maps at each EoR-Spec spectral bin. 

The TOD can be described by a linear model:
\begin{equation}
    \mathbf{d} = \mathbf{P}\,\mathbf{m} + \mathbf{n}
\end{equation}
where $\textbf{d}$ is a vector of all TOD samples, \textbf{m} is the true signal in map-space, $\textbf{P}$ is the pointing matrix that projects sky pixels onto detector samples, and $\textbf{n}$ is the noise.
Given this data model, the estimated map $\hat{\mathbf{m}}$ is obtained by solving the general map-making equation \citep{Tegmark1997_CMB_maps,  Natoli2001_PlanckMap, Patanchon_2008_SanepicMap, Dunner_2012_ACT_mapmaking}:
\begin{equation}
    \label{eq:general_map_making}
    \left( \textbf{P}^{\mathrm{T}} \textbf{N}^{-1} \textbf{P} \right) \hat{\mathbf{m}}
            = \textbf{P}^{\mathrm{T}} \textbf{N}^{-1} \mathbf{d}
\end{equation}
We assumed the noise to be Gaussian and uncorrelated between samples and detectors, such that the noise covariance matrix is diagonal:
\begin{equation}
\mathbf{N} = \langle \mathbf{n}\mathbf{n}^{\mathrm{T}} \rangle = \mathrm{diag}(1/w)
\end{equation}
where \(w\) denotes the inverse-variance weights of the TOD samples. This approximation is reasonable after PCA filtering, which largely whitens the timestreams and suppresses residual correlations. Under this assumption, the map-making equation reduces to a simple inverse-variance weighted binning of the TOD into map pixels:
\begin{equation}
\hat{\mathbf{m}}
=
\left(\mathbf{P}^{\mathsf{T}}\mathbf{W}\mathbf{P}\right)^{-1}
\mathbf{P}^{\mathsf{T}}\mathbf{W}\mathbf{d}
\end{equation}
where \(\mathbf{W} = \mathbf{N}^{-1} = \mathrm{diag}(w)\). 

A similar inverse-variance binned map-maker was used in F22, \citet{Lunde_2024_COMAP_Season2} and \citet{EChapin2013_SCUBA2iterative}. We used the \texttt{BinMap} and \texttt{MapMaker} operators within the TOAST framework to produce the binned maps. Since the data were first filtered to suppress correlated noise, leaving the TOD predominantly in the white-noise regime prior to binning, this approach is commonly referred to as the F\&B method.

\begin{figure}
  \resizebox{\hsize}{!}{\includegraphics{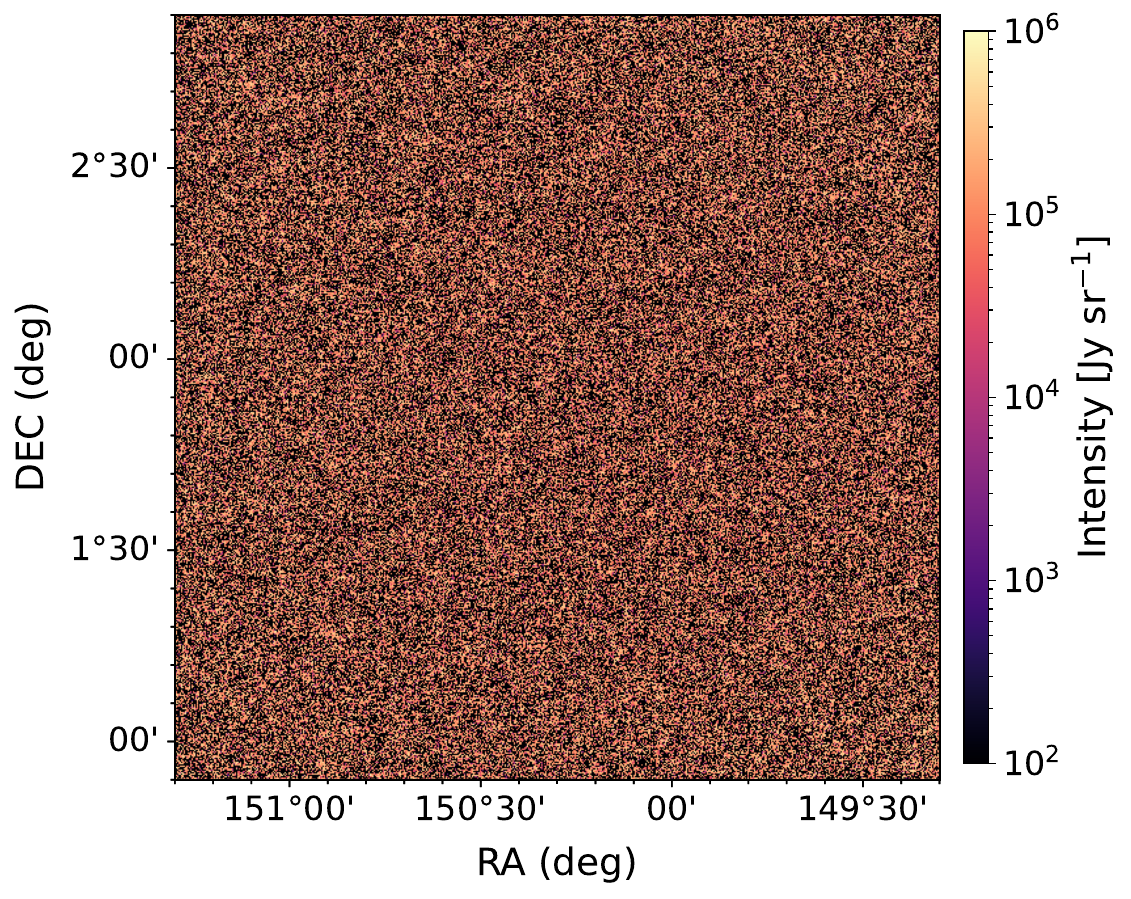}}
  \caption{A $2^\circ \times 2^\circ$ cutout region of the final filtered and binned map for the 365 GHz spectral bin, obtained after co-adding the FPI steps contributing to this bin. The resulting map is dominated by white noise. The cutout region corresponds to the green-dashed region in \Fig{fig:coadd_hits}.}
  \label{fig:cutout_fb_365}
\end{figure}

The F\&B method was used to produce mock maps for all spectral bins in the $330-370$ GHz frequency range considered in this work. As shown in \Fig{fig:dets_per_FPIchnl}, multiple FPI steps contribute to each spectral bin. We produced maps independently for each contributing FPI step and co-added the maps contributing to the same spectral bin using a noise-weighted mean:
\begin{equation}
m_p^{\mathrm{co-add}} = \frac{\sum_i w_{i,p}\, m_{i,p}}{\sum_i w_{i,p}}
\end{equation}
where \(m_{i,p}\) and \(w_{i,p}\) are the map value and inverse-variance weight of the \(i^{\mathrm{th}}\) FPI-step map at pixel \(p\), respectively. 
\Figure{fig:all_fbmaps_grid} presents the resulting co-added maps per spectral bin produced using this pipeline. We selected a $2^\circ \times 2^\circ$ sub-region within each co-added map to construct the 3-D spectral data-cube for the LIM analysis. \Figure{fig:cutout_fb_365} shows the resulting co-added F\&B map for the 4 deg\textsuperscript{2} region of the 365 GHz spectral bin. The map is mostly white-noise dominated and the correlated noise components were filtered-out by the pipeline. 

\section{Results} \label{sec:results}
\subsection{Temporal noise properties}\label{sec:temporal-noise-PSD}

\begin{figure}
  \resizebox{\hsize}{!}{\includegraphics{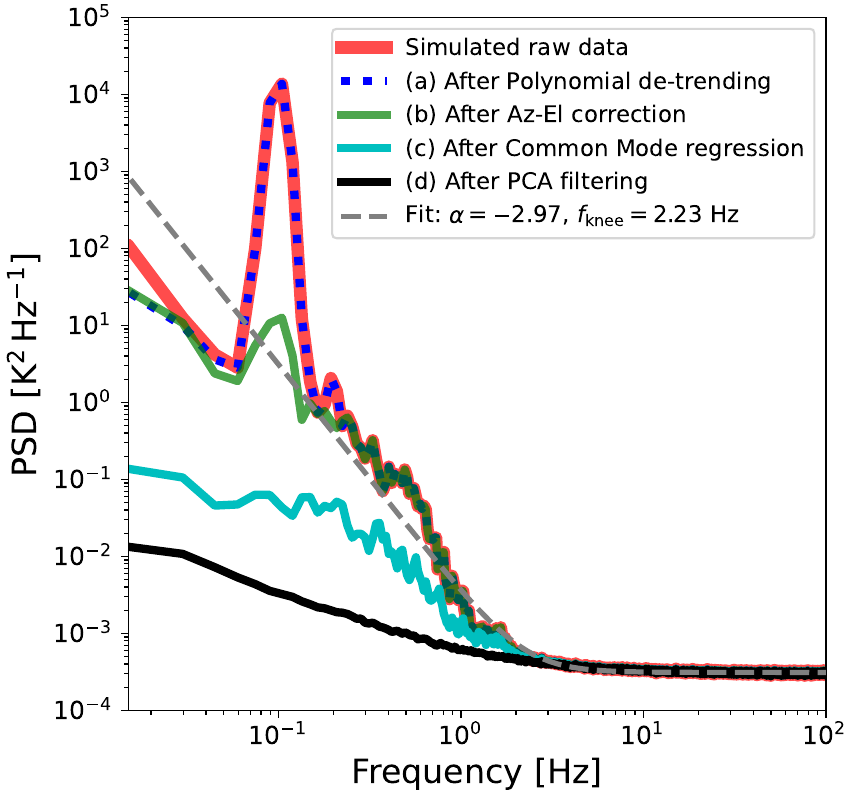}}
  \caption{Average temporal power spectral density (PSD) of the detectors in the 365\,GHz spectral bin. The red line shows the raw timestream PSD, while the dotted blue, solid green, solid cyan, and solid black lines show the PSD after polynomial de-trending, Az-El correction, common-mode regression, and PCA filtering, respectively. The dashed grey line shows the $1/f$ noise model (\Eq{eq:1_f_psd_model}) fitted to the PSD after Az-El correction, with $\alpha=-2.97$ and $f_{\mathrm{knee}}=2.23$\,Hz. The final filtered PSD is $\sim 4$ orders of magnitude lower than the raw PSD at the lowest temporal frequencies and has close to a flat-spectrum white noise behaviour.}
  \label{fig:psd_tod_f365}
\end{figure}

In this section, we characterise the noise properties of the TOD simulated in \Sect{sec:simulating-detector-timestream} and analyse the effect of the filter chain on the noise profile. The simulated TOD are dominated by correlated noise in the time domain at low temporal frequencies \footnote{The temporal frequency $f$ denotes Fourier modes in the time domain of the TOD, and is distinct from the observational frequency corresponding to the spectral bins and should not be confused.}. The correlated noise spectrum can be described using a noise power spectral density (PSD) model (F22, \citealp{Ihle2023_BeyondPlanck, Li2021_Meerkat_noise}):
\begin{equation}
    \label{eq:1_f_psd_model}
    N(f) = \sigma_w^{ 2} \left[1 + \left( \frac{f}{f_\mathrm{knee}} \right)^\alpha \right]
\end{equation}
where $\sigma_w^{ 2}$ is the white noise power, $f$ is the temporal frequency, $f_\mathrm{knee}$ is the \textit{knee frequency} at which the contributions from correlated and white noise power are equal, and $\alpha$ is a negative spectral index that determines the slope of the correlated noise spectrum. The power-law term captures the $1/f$-like behaviour of the correlated noise, with the correlated modes dominating at $f < f_\mathrm{knee}$, while the spectrum approaches flat white-noise spectrum at $f > f_\mathrm{knee}$. 

The PSD is calculated for the simulated raw TOD using the Welch method \citep{Welch1967}. We calculated the PSD for each detector corresponding to a spectral bin and an observation and averaged over the detectors. \Figure{fig:psd_tod_f365} shows the averaged PSD of the TOD at different stages of data filtering for the 365 GHz spectral bin corresponding to a simulated observation. The raw TOD PSD (plotted in red) shows a clear $1/f$-like noise behaviour, and a sharp peak seen at $\sim 0.1$ Hz corresponding to the telescope motion and turnarounds. The polynomial de-trending filter (dotted blue line) removed the global slowly varying trends that correspond to the lowest temporal frequency bins in the noise PSD. The green line shows the averaged PSD after fitting and removing the Az-El motion related systematics that targeted and removed most of the correlated noise at $\sim 0.1$ Hz.

We fitted the PSD noise model (shown in dashed grey) from \Eq{eq:1_f_psd_model} to the Az-El-corrected timestreams, for which the residual TOD is primarily dominated by the atmospheric common mode. The fit yielded $f_{\mathrm{knee}}=2.23$ Hz and $\alpha=-2.97$, close to the Kolmogorov-law power spectrum expectation of $-8/3$.
The CM-filter captured and regressed-out the common-mode structure from the TOD and the PSD after CM-filtering is shown in cyan. The CM-filter significantly flattened the $1/f$-noise slope and reduced correlated noise power by $\sim 2$ orders of magnitudes at the lowest end of the temporal frequencies. The residual correlated noise that was not picked-up by the CM-filter was subsequently removed by subtracting the first three principal components in the PCA-based filter. 
The PCA filter further reduced the power at the lowest temporal frequencies by approximately one order of magnitude. Shown in black in \Fig{fig:psd_tod_f365} is the cleaned TOD after applying the complete filter-chain. The example observation shown in \Fig{fig:psd_tod_f365} is representative of all simulated observations and spectral bins.

The cleaned TOD has a negligible correlated noise component compared to the simulated raw data, and is mostly dominated by white noise. While the cleaned PSD shows a minor residual correlated noise trend at the lowest temporal frequencies, we avoided further aggressive filtering to preserve the astrophysical signal on the largest scales which correspond to the lowest temporal frequencies. After applying the filter-chain, as described in \Sect{sec:map-making}, we used the cleaned TOD from all simulated observations and spectral bins to build the spectral maps and the tomographic data-cubes.

\subsection{LIM Power Spectrum and Sensitivity}\label{sec:lim-power-spectrum}
\begin{figure}
  \resizebox{\hsize}{!}{\includegraphics{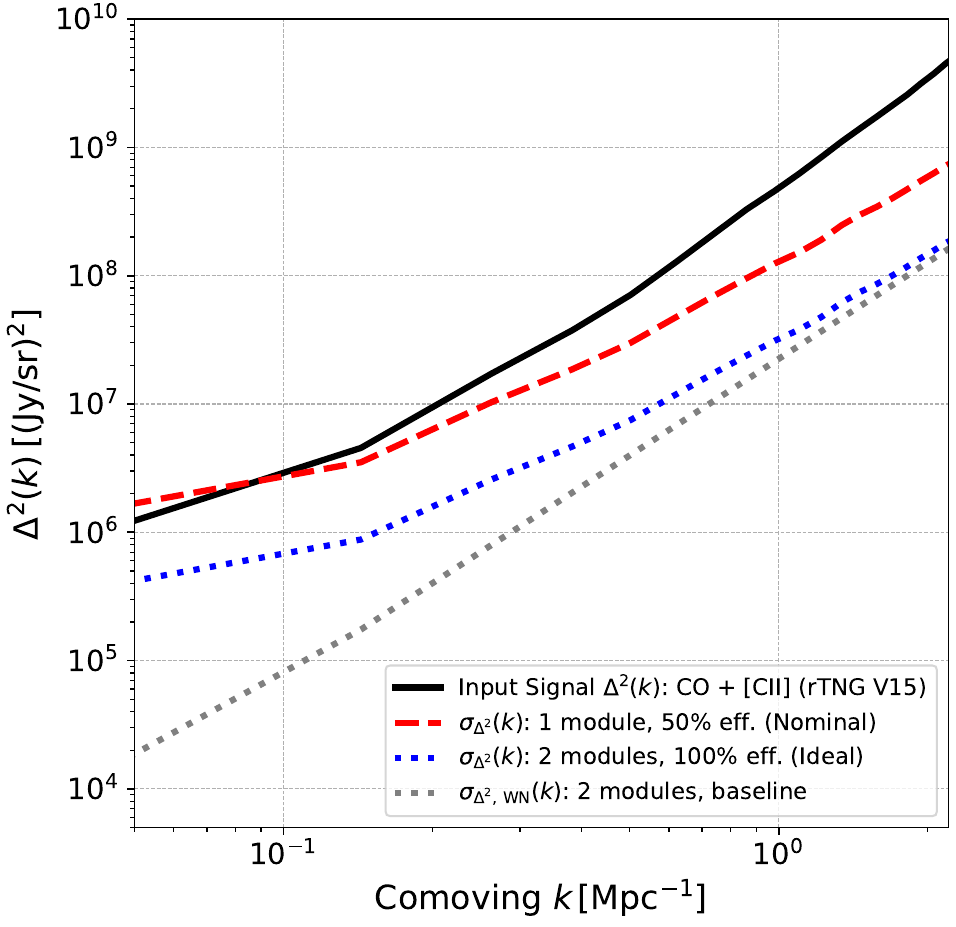}}
  \caption{Spherically averaged and normalised power spectra sensitivities for the $330 - 370$ GHz band of the EoR-Spec instrument at an average redshift of $\langle z \rangle\textsubscript{\CII} = 4.45$. The power spectra were calculated over a 4 deg\textsuperscript{2} survey area and spherically averaged with bins with $\Delta k = 0.12 \, \mathrm{Mpc}^{-1}$. Normalised power spectrum ($\Delta^{2}(k)$) for the input line-intensity signal with [C II] and CO line transitions is shown in solid black. The nominal scenario considered a single EoR-Spec module operating at 50\% observing efficiency, while the ideal scenario considered two modules observing at 100\% efficiency. $1\sigma$ sensitivities ($\sigma_{\Delta^2}(k)$) corresponding to the nominal  and ideal scenarios are shown in red-dashed and blue-dotted, respectively. The baseline white-noise only sensitivity ($\sigma_{\Delta^2, \, \mathrm{WN}}(k)$), without any signal contribution, for the ideal configuration with two EoR-Spec modules deployed for the DSS is shown for reference in dotted-grey.} 
  \label{fig:fb_ps_sens_norm}
\end{figure}

We constructed the 3-D LIM data-cubes by combining the filtered and cleaned map-level products across the $330 - 370$ GHz spectral bins, producing an intensity field $I(\mathbf{x})$ in angular coordinates and frequency. To analyse the LIM data-cube, we used the 3-D spherically-averaged power spectrum (PS) which quantifies the spatial clustering of line intensity fluctuations as a function of comoving scale. We used a method similar to K22, \citet{JClarke2024_CII_luminosity} and \citet{Chung2022_COMAP_Season1} to compute the spherically averaged PS, and we report the normalised form given by:
\begin{equation}
    \Delta^{2}(k) 
    = \frac{k^{3}}{2\pi^{2}}\,P(k)
    = \frac{k^{3}}{2\pi^{2} \, V_{\rm{box}}}
    \left\langle \lvert \tilde I(\mathbf{k}) \rvert^{2} \right\rangle_{|\mathbf{k}|\in k}
\end{equation}
where $P(k)$ is the 3-D PS of the line-intensity field, $V_{\rm{box}}$ is the co-moving survey volume, $\langle |\tilde I(\mathbf{k})|^{2} \rangle$ denotes the squared absolute value of the Fourier-transformed intensities, averaged over a spherical $k$-bin shell. The normalised spherically averaged PS, $\Delta^{2}(k)$ has units of (Jy/sr)$^{2}$.

The Fourier modes $k$ can be decomposed into components perpendicular and parallel to the line of sight, $k_{\perp}$ and $k_{\parallel}$ respectively, such that $k = \sqrt{k_{\perp}^{2} + k_{\parallel}^{2}}$. Here, $k_{\perp}$ traces transverse spatial fluctuations, while $k_{\parallel}$ probes fluctuations along the frequency (redshift) direction. We account for the effects of the angular beam and the Lorentzian spectral response, which set the effective resolutions in the transverse and line-of-sight directions, respectively. As a result, power is suppressed at large $k_{\perp}$ by the beam and at large $k_{\parallel}$ by the finite spectral resolution. The low-$k$ modes probe the clustering regime and trace large-scale structure, while the high-$k$ modes are dominated by shot-noise from the discrete distribution and luminosities of individual line-emitters.

We present the statistical uncertainty (sensitivity) of the normalised, spherically averaged PS by accounting for the number of Fourier modes, $N_m$, contributing to each $k$-bin, similar to \citet{DChung2020_Forecasting_CII} and K22, as:
\begin{equation}
    \sigma_{\Delta^2}(k)
    = \frac{k^{3}}{2\pi^{2}}\,\frac{P_\mathrm{signal + noise}(k)}{\sqrt{N_{m}(k)}}
\end{equation}
where $P_\mathrm{signal + noise}(k)$ is the spherically averaged PS of the line-intensity signal and the noise. In this work, $P_\mathrm{noise}(k)$ includes contributions from both white and correlated ($1/f$) noise components included in the simulations. The resulting PS corresponds to the simulated observed signal and noise PS rather than the intrinsic signal-only spectrum. The white-noise sensitivity can similarly be defined as:
\begin{equation}
    \sigma_{\Delta^2, \, \mathrm{WN}}(k)
    = \frac{k^{3}}{2\pi^{2}}\,\frac{P_\mathrm{WN}(k)}{\sqrt{N_{m}(k)}}
\end{equation}
where ${P_\mathrm{WN}(k)}$ is the white-noise power without any signal contribution. 

We report $1\sigma$ sensitivity ($\sigma_{\Delta^2}(k)$) estimates for nominal and ideal scenarios based on our simulations. The sensitivities are calculated for $\Delta k = 0.12\,\mathrm{Mpc}^{-1}$, identical to the power-spectrum calculation, and account for the number of Fourier modes $N_m$ in each $k$-bin. In the nominal scenario, we considered a configuration in which a single fully-populated EoR-Spec module is deployed and the LIM survey achieves an observing efficiency 
\footnote{We define the observing efficiency as the fraction of observed pixels within the target area used for the PS analysis, relative to the total number of pixels in the observed field.}
of 50\%. In the ideal scenario, we assumed that two EoR-Spec modules are deployed and that the LIM fields are observed with a 100\% observing efficiency; this scenario corresponds to the baseline configuration adopted in the \citet{CCATCollab2022_Forecasts}. Our simulations in this work considered only 50\% of the total detectors per spectral bin within a single EoR-Spec module, observing the target field with a 50\% efficiency (as stated in \Appendix{sec:scan-pattern}). The sensitivities for the nominal and ideal scenarios were obtained by appropriately scaling the noise in the simulation results. Starting from the simulated $\Delta^{2}(k)$ PS obtained from the reduced and cleaned data-cube using the filter-chain with three leading PCA components removed, we derived $1\sigma$ sensitivity estimates, accounting for the number of Fourier modes $N_\mathrm{m}$, for the nominal and ideal scenarios. For the nominal scenario, we applied a reduction factor of 2 to account for the increased number of detectors, as only 50\% of a fully populated module was simulated in this work. For the ideal scenario, we applied a reduction factor of 8, accounting for the increased number of detectors across two EoR-Spec modules and improved observing efficiency.

\Figure{fig:fb_ps_sens_norm} includes the $1\sigma$ sensitivities for the two scenarios: the dashed-red line shows the sensitivity for the case in which only a single EoR-Spec module is deployed and a $4\,\mathrm{deg}^{2}$ DSS field is observed with a 50\% observing efficiency, while the dotted-blue line shows scenario of observing the field with 100\% efficiency and deploying 2 EoR-Spec modules. The solid black curve shows the fiducial input \CII + CO PS before applying the instrumental beam and spectral response. For comparison, the $1\sigma$ white-noise sensitivity ($\sigma_{\Delta ^{2},\, \mathrm{WN}} (k)$)
\footnote{Note that $\sigma_{\Delta ^{2},\, \mathrm{WN}} (k)$ accounts only for the white-noise power and does not include the signal contribution.}
is shown as a grey-dashed line, corresponding to the white-noise reported by \citet{CCATCollab2022_Forecasts} for the 350 GHz band. We discuss the implications of the PS sensitivity results in \Sect{sec:conclusion}.

The noise term includes both instrumental white noise appropriate to each configuration and residual correlated $1/f$ noise remaining after filtering. The effects of the instrument beam and the finite spectral resolution are already folded into the simulations and are therefore accounted for in the sensitivity estimates. As each detector was simulated with its white and correlated noise components and propagated through the full data-reduction pipeline, the noise estimates presented here are realistic. The resulting simulated sensitivities are broadly consistent with the analytical sensitivity forecasts reported  
\footnote{\citet{CCATCollab2022_Forecasts} and \citet{DChung2020_Forecasting_CII} LIM sensitivity estimates used $\Delta k = 0.035 \,\mathrm{Mpc}^{-1}$.}
by \citet{CCATCollab2022_Forecasts} and \citet{DChung2020_Forecasting_CII}.

\subsection{Transfer Function Analysis and the Effect of Filtering}\label{sec:transfer-function-analysis}
The PS calculated from the processed and filtered spectral data-cube is a biased estimator of the true LIM signal, as the reduction pipeline modifies the signal through filtering, the finite resolution of the instrument, and map-making operations. In particular, large-scale modes are partially suppressed by timestream filtering and correlated-noise mitigation, leading to a scale-dependent attenuation that must be quantified and corrected using the transfer function (TF). We quantify the scale-dependent response of our end-to-end simulation pipeline using a pipeline TF. The TF, $T(k)$, is defined as the ratio between the recovered and the input signal PS as a function of wave-number $k$. 

We estimated the pipeline TF using the end-to-end simulations processed through the same filtering and map-making pipeline as the mock dataset. In the TF analysis, we focused solely on the effect of the pipeline filtering on the cosmological signal. Let $P_\mathrm{signal}(\mathbf{k})$ denote the input \CII + CO PS, including the effects of the instrumental beam and spectral response. We generated a set of simulations containing the line-intensity signal and all noise components (denoted as \textit{signal + noise}). Thereafter we produced noise-only simulations (denoted as \textit{noise}), with identical noise realisations and instrument configuration but no input line-intensity signal, and processed them through the same pipeline configuration as in \textit{signal + noise}. Under the assumption that signal and noise are uncorrelated, the noise-subtracted recovered signal power is $P_\mathrm{recovered}(k) = P_\mathrm{signal + noise}(k) \, - \, P_\mathrm{noise}(k)$. We calculated the pipeline TF as:
\begin{equation}
    T(k) = \frac{P_\mathrm{recovered}(k)}{P_\mathrm{signal}(k)}
\end{equation}
where we report $T(k)$ after applying the same spherical averaging used for the $P_\mathrm{signal + noise}(k)$ PS (described further in \Appendix{sec:effect-pca-filter-ps}) estimates.

\begin{figure}
  \resizebox{\hsize}{!}{\includegraphics{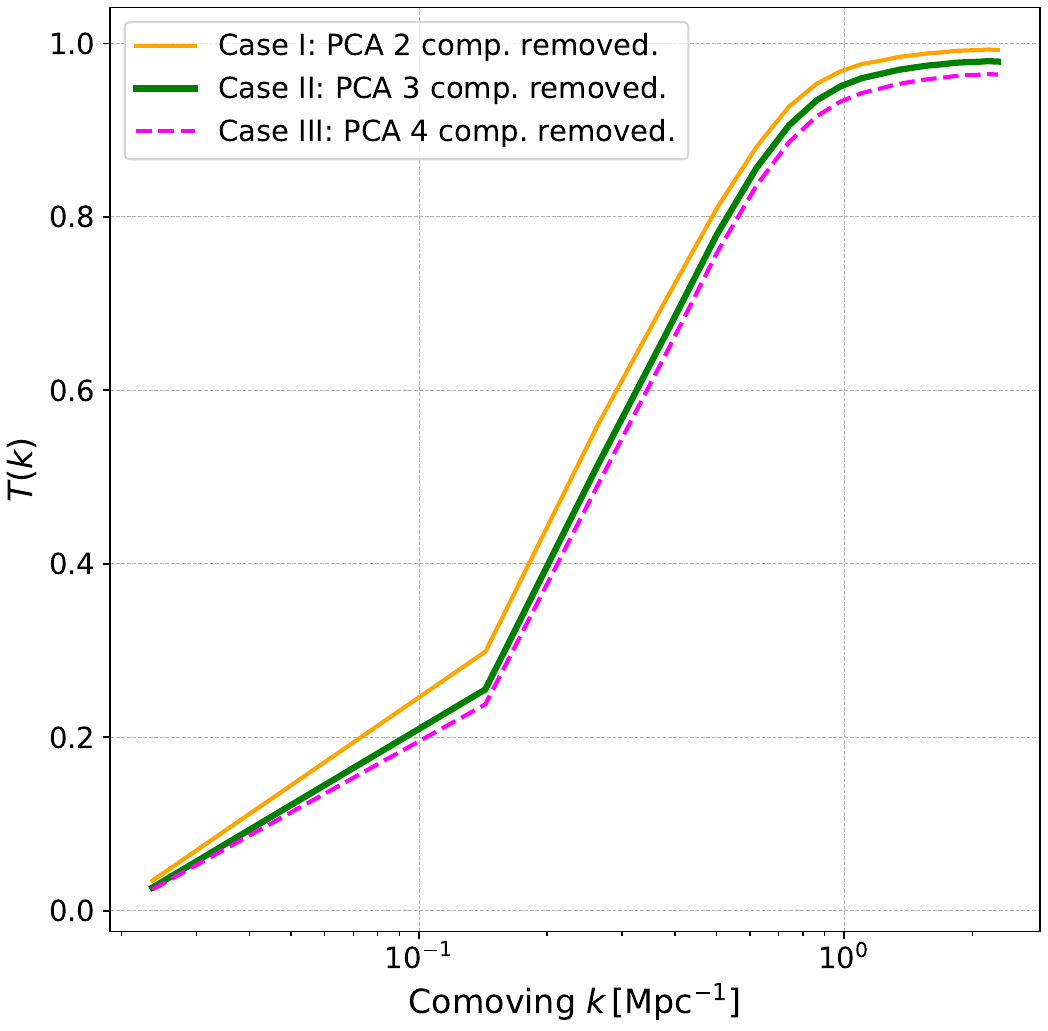}}
  \caption{Pipeline transfer function $T(k)$ for the three cases of PCA-filter with two (Case I in orange), three (Case II in green) and four (case III in dashed magenta) leading principal components subtracted respectively, after processing the data with polynomial de-trending, Az-El correction and CM regression.}
  \label{fig:TF_ps}
\end{figure}

\Figure{fig:TF_ps} shows $T(k)$ differentiating between the three cases of PCA filtering discussed in \Sect{sec:lim-power-spectrum} and \Appendix{sec:effect-pca-filter-ps}. In all three cases of pipeline filtering, we initially performed polynomial de-trending, Az-El correction and CM regression. For all three cases, $T(k)$ peaks at the smallest scales at $k > 1\,\mathrm{Mpc}^{-1}$ and is more than 80\% for $k \gtrapprox 0.5\,\mathrm{Mpc}^{-1}$. The pipeline has minimal effect in the shot-noise regime of the measured PS. $T(k)$ drops below 20\% at $k \lessapprox 0.1\,\mathrm{Mpc}^{-1}$ as the filtering significantly affects the signal recovery on the largest measured scales. The apparent corner near $k \sim 0.1\,\mathrm{Mpc}^{-1}$ is a consequence of the finite binning used in the analysis (20 uniformly spaced $k$-bins), which appears non-uniform when displayed on a logarithmic $k$ axis. The signal loss has a direct correlation with the number of leading principal components subtracted and we demonstrated that only a few PCA components should be subtracted to preserve the signal as far as possible. However, a minimum number of components must be removed to suppress excess correlated noise in the final maps. We find that, although Case I, with two PCA components removed, exhibits slightly lower signal loss, the resulting maps show excess residual $1/f$ noise. We therefore adopted Case II, with three PCA components removed, as our nominal configuration, as it yields minimal residual correlated noise (see \Fig{fig:psd_tod_f365}). Case III, in which an additional component is removed, does not provide a significant improvement in suppressing residual correlated noise and instead leads to increased attenuation of the cosmological signal. The resulting PS across all three cases agree within $\sim 5\%$, indicating that the differences between the case are relatively modest.

\section{Summary and Conclusion}\label{sec:conclusion}
In this study, we developed an end-to-end simulation framework for EoR-Spec to investigate the impact of atmospheric $1/f$ noise and data-reduction filtering on recovery of the \CII + CO LIM PS. Using realistic mock observations spanning $330 - 370$ GHz, we produced LIM data-cubes and PS with a F\&B pipeline and quantified the corresponding transfer function. Our results show that the pipeline effectively suppresses atmospheric contamination while preserving the underlying LIM signal, although it attenuates power on the largest spatial scales. We discuss the implications of these results for future EoR-Spec observations and the current limitations of the analysis.

The sensitivity forecasts indicate that the DSS can place meaningful constraints on the \CII + CO PS at $\langle z \rangle\textsubscript{\CII} = 4.45$. In the nominal scenario, corresponding to a fully populated EoR-Spec module with 50\% observing efficiency, the survey can detect the fiducial signal for $k > 0.1\,\mathrm{Mpc}^{-1}$ using $\Delta k = 0.12\,\mathrm{Mpc}^{-1}$ bins. With two modules and maximum observing efficiency, the sensitivity extends across the full range considered here, $0.02\,\mathrm{Mpc}^{-1} < k < 2\,\mathrm{Mpc}^{-1}$. Our forecasts show that an observing efficiency $\gtrsim 50\%$ and the deployment of a second EoR-Spec module significantly improve the S/N of the \CII + CO PS measurement, enabling detections across all probed $k$-modes.

We adopted the fiducial \CII and CO models from K22 and K24 as input signal models for the simulations and sensitivity forecasts. Accounting for atmospheric $1/f$ noise and observing efficiency, the nominal DSS sensitivity is sufficient to detect the signal assuming the fiducial models and is expected to provide even stronger constraints for more optimistic scenarios with brighter \CII emission (see Fig.~6 of K22 for a comparison of the different \CII models).

The filtering chain substantially suppresses correlated low-frequency noise across the $330-370$ GHz band, but at the cost of the large-scale clustering modes. The TF remains above 80\% at $k \gtrapprox 0.5\,\mathrm{Mpc}^{-1}$ and falls below 20\% at $k \lessapprox 0.1\,\mathrm{Mpc}^{-1}$. As a result, modes with $k \gtrapprox 0.5\,\mathrm{Mpc}^{-1}$ remain reasonably well constrained after correction, whereas recovery becomes increasingly difficult below $k \lessapprox 0.1\,\mathrm{Mpc}^{-1}$. The pipeline is therefore well suited to the shot-noise regime but less effective for measuring large-scale clustering. This behaviour is expected because atmospheric contamination primarily overlaps with the largest angular and spectral modes of the LIM signal. 

We also investigated the impact of removing the leading PCA components from the TOD. The optimal number of components to be removed should not be regarded as universal, but instead re-evaluated for each dataset because additional correlations and systematics may be present in real observations. The methodology demonstrated here nevertheless provides a practical framework for future EoR-Spec data reduction. In addition to the pipeline TF analysed here, the instrumental beam suppresses power on small transverse scales, while the Lorentzian spectral response affects line-of-sight modes \citep{Lunde_2024_COMAP_Season2, Marcuzzo2025_CII_LIM}. 

Several simplifying assumptions were adopted in this study. We did not include gain fluctuations, ground pickup, data glitches, or continuum foregrounds such as the cosmic microwave background and cosmic infrared background. We also did not differentiate between the \CII LIM signal and the CO line interlopers in this analysis. We refer the reader to \citet{Clarke2026}, which investigates the recovery of the \CII LIM signal in the presence of CO foregrounds using correlated noise simulations derived from this work. Furthermore, we assumed an idealised EoR-Spec instrument model with perfect knowledge of the detector spectral response and spectral binning. Effects such as non-ideal bandpass filters, spectral cross-talk, and spillover between neighbouring bins were not included and may introduce additional mode mixing or reduce sensitivity. These effects should be investigated using measured instrument performance as EoR-Spec commissioning progresses. 

Future work should explore map-making approaches aimed at recovering larger-scale modes, including maximum-likelihood, generalised least-squares, and destriping methods \citep[e.g.][]{Zhang2025_21cmLIM_maps, Keihanen_2010_madam, Dunner_2012_ACT_mapmaking, Naess_2025_ACT_DR6map}. Overall, the F\&B pipeline provides a well-characterised baseline for early DSS analyses and is particularly effective for the shot-noise regime. As EoR-Spec begins observations, it can serve as a starting point for the first \CII + CO LIM measurements while more advanced map-making methods are developed. 

The main findings of this study are summarised below:
\begin{enumerate}
    \item We demonstrated that an optimised data-reduction pipeline can effectively suppress correlated noise across the $330-370$ GHz band. After polynomial de-trending, scan-synchronous corrections, CM subtraction, and removing the first 3 principal components, the residual power spectral density is largely consistent with white noise. The pipeline reduced the correlated low-frequency component by $\sim 4$ orders of magnitude at the lowest temporal frequencies, bringing the residual spectrum close to a flat, white-noise-like regime (see \Fig{fig:2d_tod_clean} and \Fig{fig:psd_tod_f365}).
    \item  In a nominal scenario, where a single EoR-Spec module is used for the DSS at 50\% observing efficiency, the survey achieves sufficient sensitivity to detect the \CII + CO PS at a mean redshift of $\langle z \rangle\textsubscript{\CII} = 4.45$ on scales $k > 0.1\,\mathrm{Mpc}^{-1}$, considering uniform $k$-bin of width $\Delta k = 0.12 \, \mathrm{Mpc}^{-1}$. In the ideal scenario, with maximum observing efficiency and two EoR-Spec modules, the instrument will be capable of measuring the \CII + CO PS across all probed $k$-scales ($0.02\,\mathrm{Mpc}^{-1} < k < 2\,\mathrm{Mpc}^{-1}$). Both forecasts assume the fiducial TNG300 and V15-based [C\,{\sc ii}] and CO signal models adopted in this work, and account for realistic instrument noise and residual correlated noise following pipeline filtering (see \Fig{fig:fb_ps_sens_norm}).
    \item  The pipeline transfer function, $T(k)$, of the filter-chain is greater than 80\% at $k \gtrapprox 0.5\,\mathrm{Mpc}^{-1}$ and drops below 20\% at $k \lessapprox 0.1\,\mathrm{Mpc}^{-1}$, showing substantial signal loss on large scales due to filtering. The shot-noise regime at $k > 0.5\,\mathrm{Mpc}^{-1}$ remains largely unaffected by the filtering (see \Fig{fig:TF_ps}). Recovering large-scale clustering modes remains challenging with a filtering-based data-reduction pipeline.
    \item Using the fiducial \CII and CO models from K22 and K24, we find that the DSS sensitivity should be sufficient to detect the signal in the shot-noise dominated regime in both the nominal and ideal survey configurations, with stronger sensitivity expected for more optimistic scenarios with brighter \CII emission.
    \item The observing efficiency and the detector sampling per spectral bin are key survey-design parameters that set the effective depth of the spectral maps and therefore the final PS sensitivity. Maintaining high observing efficiency ($\gtrsim 50\%$) and maximising the number of detectors per spectral bin are critical to placing meaningful constraints on the \CII and CO PS measurements.

\end{enumerate}

\begin{acknowledgements}
We warmly thank Theodore Kisner and Reijo Keskitalo for their invaluable support in utilising the TOAST software framework and for their continued efforts in maintaining and improving the software repository. We gratefully acknowledge the access to the Marvin HPC cluster of the University of Bonn and the support provided by the HPC@HRZ Team of the University of Bonn. We also thank Gordon Stacey, Nicholas Battaglia, Michael Niemack, Steve Choi, Zachary Huber, Ben Keller and other members of the CCAT Collaboration for their support and feedback. We are grateful to Patrick Breysse, Guilaine Lagache, Eiichiro Komatsu, Benjamin Magnelli, Cristiano Porciani, Emilio Romano-Diaz, Elena Marcuzzo, Anirban Roy, Vyoma Muralidhara, Toma B\u{a}descu, Ralf Antonius Timmermann, Reinhold Schaaf, Maude Charmetant, Stefanie M\"{u}hle, Sylvia Adscheid, Karolina Garcia, Jonas Lunde, Nils-Ole Stutzer for valuable comments at various stages of this work.
This work was supported by the Collaborative Research Centre SFB 1601 (project ID 500700252), sub-project C3, funded by the Deutsche Forschungsgemeinschaft (DFG, German Research Foundation). We acknowledge the support of the International Max Planck Research School for Astronomy and Astrophysics (IMPRS A\&A) at the Universities of Bonn and Cologne.
The CCAT project, FYST and Prime-Cam instrument have been supported by generous contributions from the Fred M. Young, Jr. Charitable Trust, Cornell University, and the Canada Foundation for Innovation and the Provinces of Ontario, Alberta, and British Columbia. The construction of the FYST telescope was supported by the Gro{\ss}ger{\"a}te-Programm of the German Science Foundation (DFG) under grant INST 216/733-1 FUGG, as well as funding from Universit{\"a}t zu K{\"o}ln, Universit{\"a}t Bonn, the Max Planck Institut f{\"u}r Astrophysik, Garching, and Duke University. The construction of EoR-Spec is supported by NSF grant AST-2009767. 
The simulations and data analysis presented in this work made use of \texttt{TOAST}, \texttt{numpy}, \texttt{scipy}, \texttt{astropy}, and \texttt{matplotlib}.
The flowchart presented in this paper was created using \texttt{Lucidchart} (\url{https://lucid.app/}).
\end{acknowledgements}

\bibliographystyle{aa_url} 

\bibliography{adev_etal_aanda2026}

\begin{appendix}

\section{Fabry-P\'erot Interferometer Theory}\label{sec:fpi-theory}
Maximum transmission in a Fabry-P\'erot Interferometer (FPI) occurs under the resonant condition, when the emergent beams are in phase and constructively interfere. Resonance happens when the optical path difference between successive internal reflections equals an integer multiple of the wavelength, satisfying the condition:
\begin{align}
	m &= \frac{2nd \cos \theta}{\lambda_0}
\end{align}
where m, the \textit{order of interference}, takes integer values, $d$ is the cavity gap thickness between the parallel reflective mirrors, $\lambda_0$ is the wavelength in vacuum, $n$  is the refractive index of the medium inside the cavity, and $\theta$ is the angle of incidence relative to the normal. The phase offset $\delta$ between two consecutive transmitted beams is expressed as:
\begin{align}
	\delta           &= 2 \pi m = \frac{4 \pi}{\lambda_0} nd \cos \theta \\
    \frac{\delta}{2} &= \frac{\pi \nu_0}{{\Delta \nu}_{\mathrm{FSR}}}
\end{align}
where the \textit{free spectral range} (${\Delta \nu}_{\mathrm{FSR}}$), when described in frequency terms, is given by
\begin{align}
    {\Delta \nu}_{\mathrm{FSR}} = \frac{c}{2 nd \cos \theta}
\end{align}
and $\nu_0$ is the $m\textsuperscript{th}$ order resonance frequency of the etalon.

In general, for multi-beam interference, the transmitted intensity distribution is described by the Airy function, which quantifies the fraction of incident light transmitted through the cavity as a function of the phase difference between successive beams. In an ideal, lossless Fabry-P\'erot system, the sum of the transmitted and reflected intensity coefficients satisfies the relation $ \mathcal{T} + \mathcal{R} = 1$, where $\mathcal{T}$ is the transmittance and $\mathcal{R}$ is the reflectance of the mirrors. This condition assumes that there is no absorption or scattering within the cavity. For an ideal Fabry-P\'erot etalon, the ratio of transmitted intensity ($I_t$) to incident intensity ($I_i$), i.e. the transmittance $\mathcal{T}$ is given by:
\begin{equation}
	\label{eq:transmission1_fpi}
	\mathcal{T} = \frac{I_t}{I_i} = \frac{1}{1+ F \sin^2 \left(\frac{\delta}{2} \right)}
\end{equation}
wherein the parameter $F$ is $\frac{4\mathcal{R}}{(1-\mathcal{R})^2}$. $F$ characterizes the spectral selectivity of the interferometer, with higher values providing more sharply defined and narrower transmission peaks at resonance. 

Hereafter, $\nu$ denotes the optical frequency of the incident radiation, and we define $\Delta \nu = \nu - \nu_0$ as its frequency offset from the resonance $\nu_0$. \Equation{eq:transmission1_fpi} can be re-written in terms of ${\Delta \nu}_{\mathrm{FSR}}$ as \Eq{eq:transmission2_fpi}. Transmission maxima occur when the resonance condition is met, resulting in evenly spaced peaks separated by the free spectral range (${\Delta \nu}_{\mathrm{FSR}}$). Near any resonance peak, we may use the small-angle approximation, 
$\sin \left(\frac{\pi \,\Delta \nu}{{\Delta \nu}_{\mathrm{FSR}}}\right) \approxeq \left(\frac{\pi \,\Delta \nu}{{\Delta \nu}_{\mathrm{FSR}}}\right) $. 
So, \Eq{eq:transmission2_fpi} can be expressed as:
\begin{align}
	\label{eq:transmission3_fpi}
	\mathcal{T}(\nu \,; d,\theta) 
                &= \frac{1}{1 + \frac{4 \mathcal{R}}{(1 - \mathcal{R})^2}  \left(  \frac{\pi \,\Delta \nu}{{\Delta \nu}_{\mathrm{FSR}}} \right)^{2}} \nonumber \\
				&= \frac{1}{1 + \left( \frac{2 \,\Delta \nu}{\Gamma}\right)^2} 
				 = \frac{\left(\frac{\Gamma}{2}\right)^2} {\left(\frac{\Gamma}{2}\right)^2 + (\Delta \nu)^{2}} \\
    \text{where } \; \Gamma &= \frac{(1- \mathcal{R})\; {\Delta \nu}_{\mathrm{FSR}} }{\pi \sqrt{\mathcal{R}}}	\nonumber
\end{align}
\Equation{eq:transmission3_fpi} represents the ideal transmission profile of a single, individual resonance profile of the FPI in the form of a normalized Lorentzian function, with full-width at half-maximum (FWHM) given by:
\begin{equation}
    \label{eq:finesse}
    \Delta \nu_\mathrm{FWHM} = \Gamma = \frac{{\Delta \nu}_{\mathrm{FSR}}}{\mathcal{F}}
\end{equation}
wherein $\mathcal{F} = \frac{\pi \sqrt{\mathcal{R}}}{1-\mathcal{R}} = \frac{\pi \sqrt{F}}{2}$ is defined as the \textit{finesse} for the spectrometer. $\mathcal{F}$ quantifies how sharply an interferometer discriminates neighbouring resonances. The \textit{resolving power} $R$ of the spectrometer is defined as $R = \frac{\nu}{\Delta \nu_\mathrm{FWHM}}$.

\section{Distribution of Detector Counts}\label{sec:detector-counts}
\begin{figure*}[!t]
\centering
  \includegraphics[width=17cm]{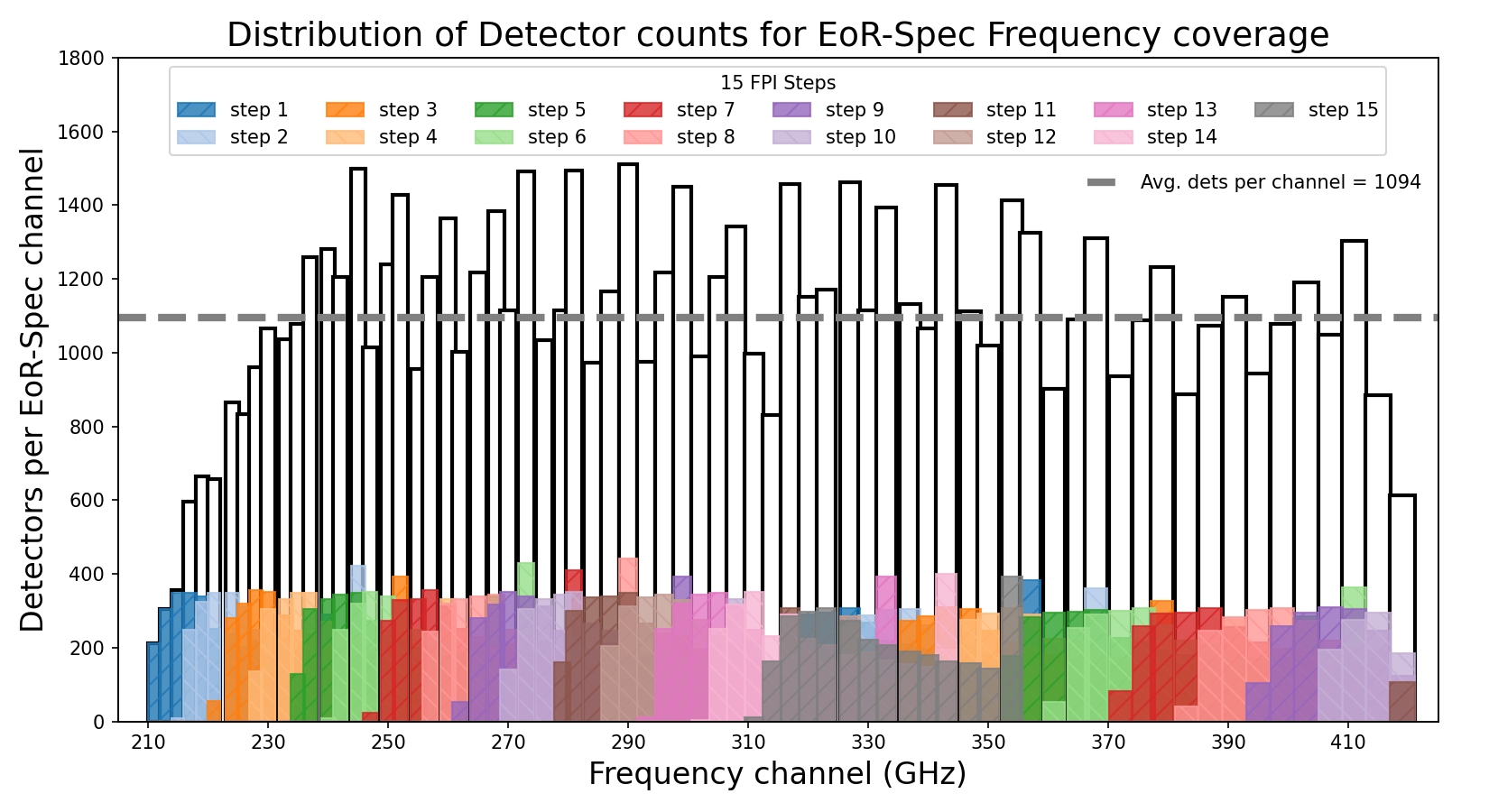}
  \caption{Distribution of detector counts per EoR-Spec spectral bin across 15 FPI steps. Each detector was assigned to a bin according to its respective peak transmitted resonance frequency. Individual colours corresponding to the 15 steps represents the contribution of detectors to each bin from a single FPI step.  The overlaid black outline shows the cumulative number of detectors per spectral bin, and the dashed gray line marks the average detector count per spectral bin.}
  \label{fig:dets_per_FPIchnl}
\end{figure*}
\Figure{fig:dets_per_FPIchnl} 
    \footnote{This figure was adapted from earlier work by Mahiro Abe and Thomas Nikola. We gratefully acknowledge their contributions to its initial development.}
illustrates how the number of detectors (or detector counts) per spectral bin in each of the 15 FPI steps contribute to the overall spectral coverage of the EoR-Spec instrument. Although each FPI step samples only a subset of the spectral bins, their combined contributions ensure uniform coverage across the full band. The 15 FPI steps were chosen to provide a fairly uniform distribution of detector counts per spectral bin after completing the full scan sequence. Each spectral bin is observed multiple times across different FPI steps, resulting in well-sampled and redundant coverage in frequency space.

\section{Mock Observation Strategy}\label{sec:scan-pattern}
\begin{figure}[!h]
  \resizebox{\hsize}{!}{\includegraphics{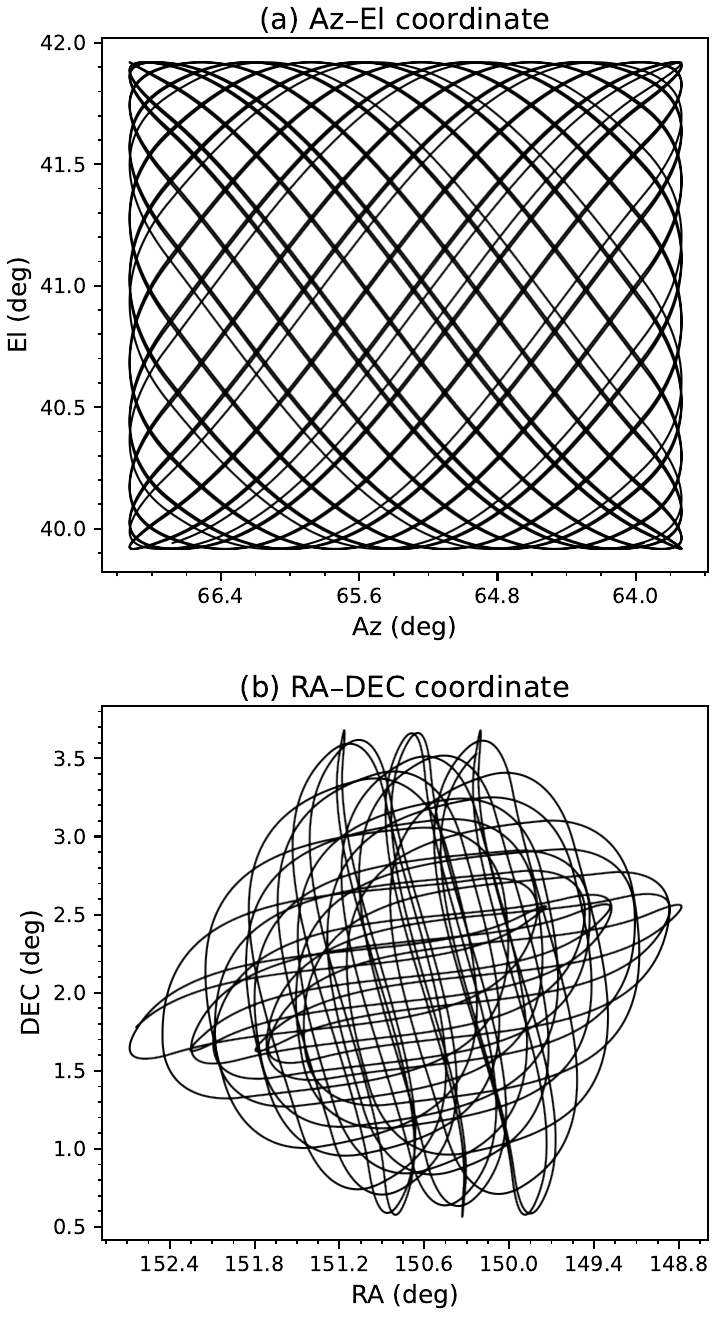}}
  \caption{The elevation-modulated azimuth scan pattern for a single mock observation of $\sim 5$ minutes is shown in Az-El coordinates (Panel a), with the corresponding pattern projected onto RA-DEC coordinates is shown in Panel (b). This scan strategy avoids sharp edges at the turnarounds.}
  \label{fig:scan_path}
\end{figure}

\begin{figure}[!h]
  \resizebox{\hsize}{!}{\includegraphics{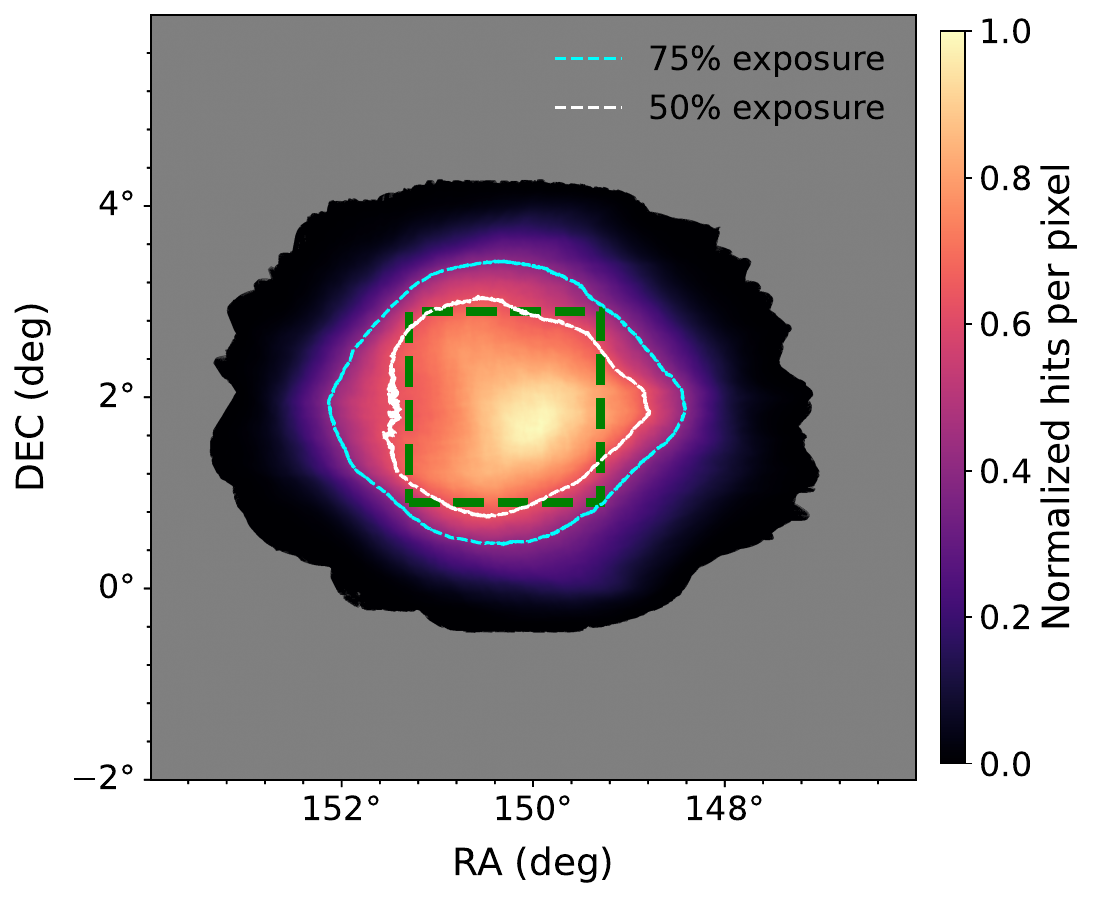}}
  \caption{Normalized exposure map of the E-COSMOS field for the 365 GHz spectral bin using the elevation-modulated azimuth scan pattern. The white dashed line marks the $50\%$ contour of the total observed pixels, while the cyan dashed line indicates the $75\%$ contour. The green dashed box shows the $2^{\circ} \times 2^{\circ}$ area selected for analysis in this work.}
  \label{fig:coadd_hits}
\end{figure}

The DSS fields targeted by the LIM survey with FYST cover an area of $2^\circ \times 2^\circ$ each, which are comparatively compact fields in contrast to the wider-area targets used in cosmological surveys like Atacama Cosmology Telescope (ACT) or Simons Ovservatory (SO) \citep{ACTcollab2016_Surveystrategy,HEbina2022_widefield,Stevens2018_Designsnextgen}. This smaller field size enables extremely deep integrations to be achieved within the nominal five-year EoR-Spec survey. Each instrument module in Prime-Cam, including the spectrometer, has a field of view of approximately $1.3^\circ$ \citep{CCATCollab2022_Forecasts}, which is comparable to the full DSS field size. For a nominal on-sky scan speed of $0.5 \,\mathrm{deg/s}$, as chosen for the simulations in this study, the telescope must reverse scan direction within a few seconds to keep the scan path confined within the compact field area. Increasing the on-sky footprint beyond this would reduce efficiency, as more time would be spent scanning outside the science field. However, practical constraints complicate the choice of scan strategy. These include the mechanical acceleration limits of the telescope in azimuth and elevation, the need to avoid large jerks at turnaround points, and the requirement for sufficiently high scan speeds to mitigate atmospheric noise. The optimal scan pattern chosen for the DSS LIM experiment with FYST will ultimately depend upon the on-sky performance observed during commissioning of the telescope.

For the simulated observations in this work, we scanned the E-COSMOS field using a elevation-modulated azimuth scan pattern. This choice allowed us to study a compact scan pattern, while providing cross-linking and sky coverage comparable to a `pong' scan developed for the SCUBA-2 instrument on the James Clerk Maxwell Telescope \citep{HThomas2014_SCUBA2Data}. The data-reduction pipeline described in \Sect{sec:data-reduction-pipeline} is equally applicable to other scan strategies, including Lissajous, `pong' and constant-elevation scans.

We used the TOAST \texttt{SimGround} operator to simulate drift-scan observations of the target field as it moved through the observed scan region. The azimuth range was kept fixed during each sweep and was not continuously adjusted to track the field centre. After each sweep, the elevation was incremented slightly, and the azimuth limits were recomputed to follow the field as it drifted on the sky. The azimuthal scan rate was set to a constant rate of $0.5 \, \mathrm{deg/s}$ on the sky, while the azimuth and elevation acceleration were  set to $1 \, \mathrm{deg/s^2}$. To simulate change in elevation, the telescope elevation was modulated sinusoidally with an amplitude of $1$ deg at a frequency of $0.1$ Hz in the Az-El frame. The elevation scanning rate was set to $1.5 \, \mathrm{deg/s}$. This elevation-modulated scan configuration provided cross-linking between successive scans while avoiding large elevation changes within a sweep to keep the scan pattern compact. \Fig{fig:scan_path} shows the resulting on-sky scan pattern in the Az-El and RA-DEC coordinates for a mock observation of $\sim 5$ minutes. We achieved a compact scan pattern using the elevation-modulated azimuth scan by limiting the azimuth scan range to approximately match the elevation scan amplitude.

\Figure{fig:coadd_hits} shows the total normalised hit count per pixel for the 365 GHz spectral bin, obtained using the elevation-modulated azimuth scan pattern described above, after combining all simulated scans contributing to this bin. For our analysis in this work, we selected a $2^{\circ} \times 2^{\circ}$ sub-region encompassing the area with approximately $50\%$ of the total exposure, corresponding to the deepest part of the map. This cutout yields a scan efficiency (or observing efficiency) of roughly $50\%$, translating to an on-field integration time of 67 hours per FPI step. 

\section{Atmospheric simulation parameters}\label{sec:atm-sim-params}
The atmospheric emission was simulated using the \texttt{SimAtmosphere} operator within the TOAST framework. The simulation employed a 3-D Kolmogorov turbulence model in which the atmosphere was represented by a moving volume of turbulent water-vapour fluctuations \citep{Errard2015_Atm_model, Morris2022_ACT}. Two atmospheric components were combined to model turbulence on different physical scales. The coarse component described large-scale fluctuations that produce slowly varying atmospheric drifts and correlated $1/f$ noise, while the fine component modelled small-scale fluctuations and short-timescale variations. The principal parameters used in the atmospheric simulations are listed in Table~\ref{tab:atm_params}.

The parameters $l_\mathrm{min}$ and $l_\mathrm{max}$ correspond to the dissipation and injection scales of the Kolmogorov turbulence model, respectively. The quantities $x_\mathrm{step}$, $y_\mathrm{step}$, and $z_\mathrm{step}$ define the size of the simulated atmospheric volume elements, while $z_\mathrm{max}$ sets the maximum height of the turbulent water-vapour volume above the telescope that is included in the atmospheric simulation. The parameter \texttt{wind\_dist} defines the maximum horizontal drift of the atmospheric volume before a new realisation is generated. The \texttt{gain} parameter rescales the simulated atmospheric TOD, and an elevation-dependent atmospheric loading is included.

\begin{table*}
\caption{Parameters used for the atmospheric simulations. Two atmospheric components were combined to model large-scale and small-scale turbulence.}
\label{tab:atm_params}
\centering
\begin{tabular}{lcc}
\hline\hline
Parameter & Coarse component & Fine component \\
\hline
Dissipation scale, $l_\mathrm{min}$ [m] & $300 \pm 30$ & $0.001 \pm 0.0001$ \\
Injection scale, $l_\mathrm{max}$ [m] & $10\,000 \pm 1000$ & $1.0 \pm 0.1$ \\
Grid spacing, $(x,y,z)$ [m] & $(50,50,50)$ & $(4,4,4)$ \\
Maximum integration height, $z_\mathrm{max}$ [m] & 2000 & 200 \\
Maximum wind drift, \texttt{wind\_dist} [m] & 10\,000 & 1000 \\
TOD scaling factor, \texttt{gain} & $1\times10^{-4}$ & $4\times10^{-5}$ \\
Field of view [deg] & 1.5 & 1.5 \\
\hline
\end{tabular}
\end{table*}

\section{Effect of PCA filtering on LIM Power Spectrum}\label{sec:effect-pca-filter-ps}
\begin{figure}
  \resizebox{\hsize}{!}{\includegraphics{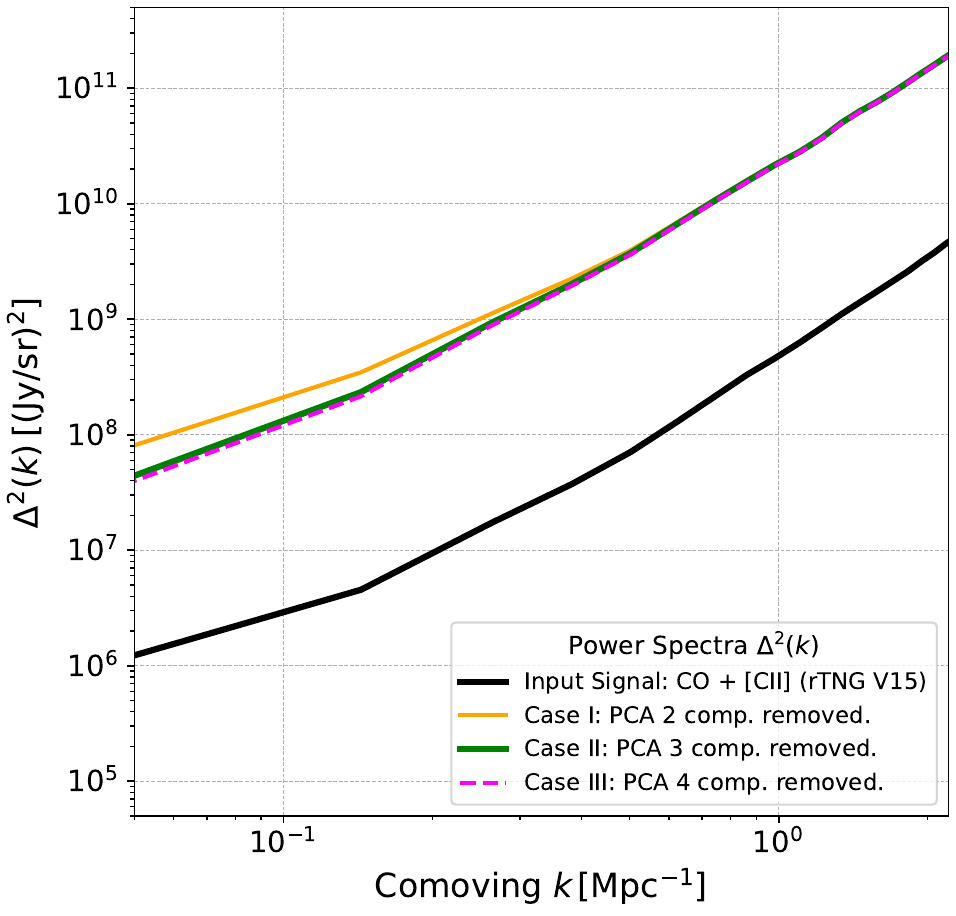}}
  \caption{Spherically averaged and normalised PS for the $330 - 370$ GHz band of the EoR-Spec instrument at an average redshift of $\langle z \rangle\textsubscript{\CII} = 4.45$. The PS were calculated over a 4 deg\textsuperscript{2} survey area and spherically averaged with bins with $\Delta k = 0.12 \, \mathrm{Mpc}^{-1}$. Normalised PS $\Delta^{2}(k)$ for the fiducial input line-intensity signal (solid black) with [C II] and CO line transitions and for the three PCA filtering cases in the data-reduction pipeline: Case I with two principal components removed (orange), Case II with three leading principal components removed (green), and Case III with four components removed (dashed magenta).} 
  \label{fig:fb_ps_norm}
\end{figure}

We explored three filtering cases for the PS (normalised $P_\mathrm{signal + noise}(k)$) analysis. In the data reduction pipeline, after performing polynomial de-trending, correcting for telescope motion, and applying CM regression in all cases, we performed PCA filtering with different numbers of components removed: (i) Case I, in which two leading PCA components were removed; (ii) Case II, with three PCA components removed; and (iii) Case III, with four PCA components removed. These cases were tested to assess the impact of progressively removing additional components on the recoverable LIM PS and to identify the optimal number of components for the mock dataset considered in this work. We analysed these cases in the context of the transfer-function analysis in \Sect{sec:transfer-function-analysis}. Case II, with three PCA components removed, was adopted as our nominal configuration and is described in \Sect{sec:pca-filter}. 

The spherically averaged and normalised PS, $\Delta^{2}_\mathrm{signal + noise}(k)$, corresponding to the three PCA filtering cases are shown in \Fig{fig:fb_ps_norm} for the $330-370$ GHz spectral range, corresponding to a \CII-emitter mean redshift of $\langle z \rangle\textsubscript{\CII} = 4.45$, and are plotted using a uniform bin width of $\Delta k = 0.12\,\mathrm{Mpc}^{-1}$. These PS correspond to the survey area of $2^\circ \times 2^\circ$ considered in this work and to the simulated configuration in which 50\% of the detectors in each spectral bin of a single EoR-Spec module were included and an on-sky observing efficiency of 50\% was achieved. The solid black line corresponds to the PS of the fiducial input signal, comprising the \CII and CO line emission described in \Sect{sec:input-signal-maps}, before convolution with the instrument beam and spectral response. Case I (solid orange), with only two PCA components removed, exhibits excessive residual $1/f$ noise that contaminates the large-scale modes around $k \simeq 0.1\,\mathrm{Mpc}^{-1}$. Case II (solid green), with three PCA components removed, reduces the excess correlated noise that is not fully captured by the first two principal components. Case III (dashed magenta), in which an additional component is removed, shows no clear improvement in correlated-noise suppression and instead risks attenuating the underlying cosmological signal through more aggressive filtering. We adopt the removal of the first three PCA components as the optimal data-reduction set-up and use it as our nominal case for the sensitivity estimates.

\section{Simulation Flowchart}\label{sec:simulation-flowchart}
\begin{figure*}
\centering
  \includegraphics[width=17cm]{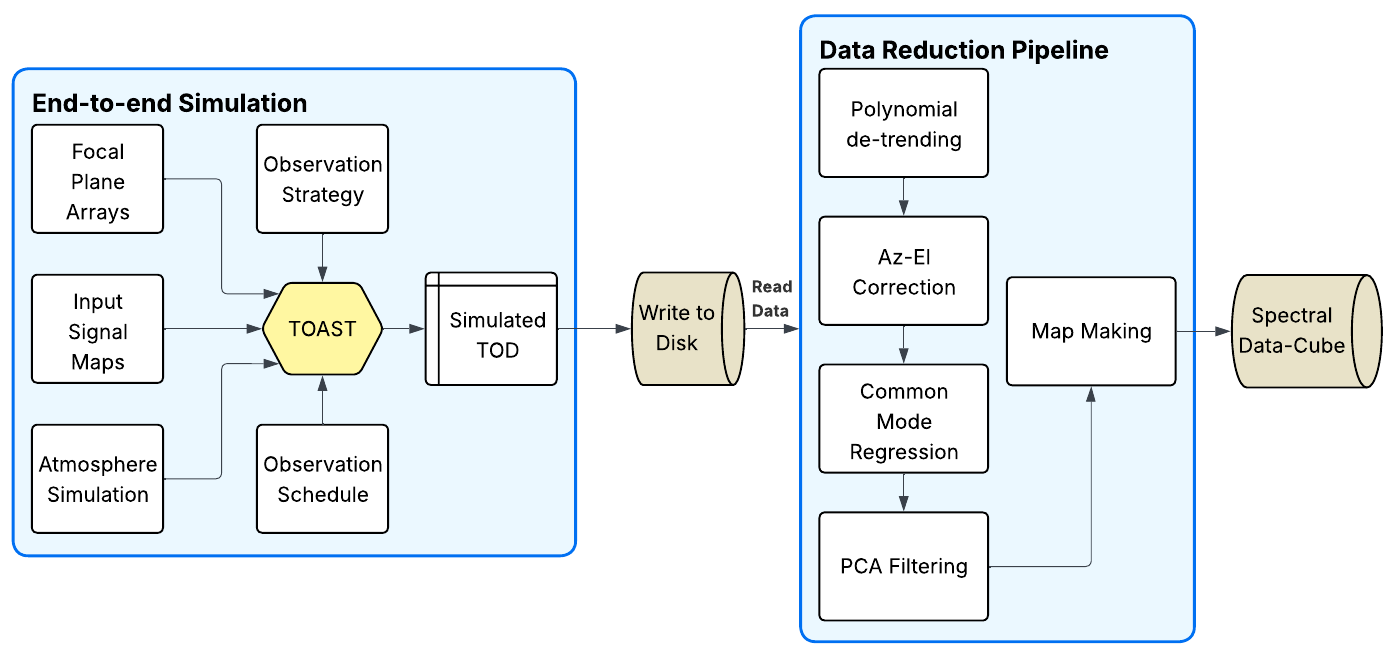}
  \caption{Flowchart showing the end-to-end simulation framework. The TOAST simulation suite uses simulated focal plane arrays, input astrophysical signal maps, simulated atmosphere, telescope observation strategy and schedules to produce the TOD which is written to disk. The data reduction pipeline is used to clean the correlated noise and produce spectral maps for LIM analysis using the mock data. }
  \label{fig:eorspec_data_flowchart}
\end{figure*}

\Figure{fig:eorspec_data_flowchart} summarises the full end-to-end workflow used in this study. In the simulation stage, the focal-plane arrays, input signal maps, atmospheric model, observing strategy, and observation schedule were passed to TOAST to generate mock TOD. The simulated TOD for all simulated spectral bins were written to disk and then read by the reduction pipeline, where polynomial de-trending, azimuth-elevation correction, common-mode regression, and PCA filtering were applied before map-making and the construction of the final LIM spectral data-cube.

\section{Timestream Filtering}
\begin{figure*}
\centering
  \includegraphics[width=17cm]{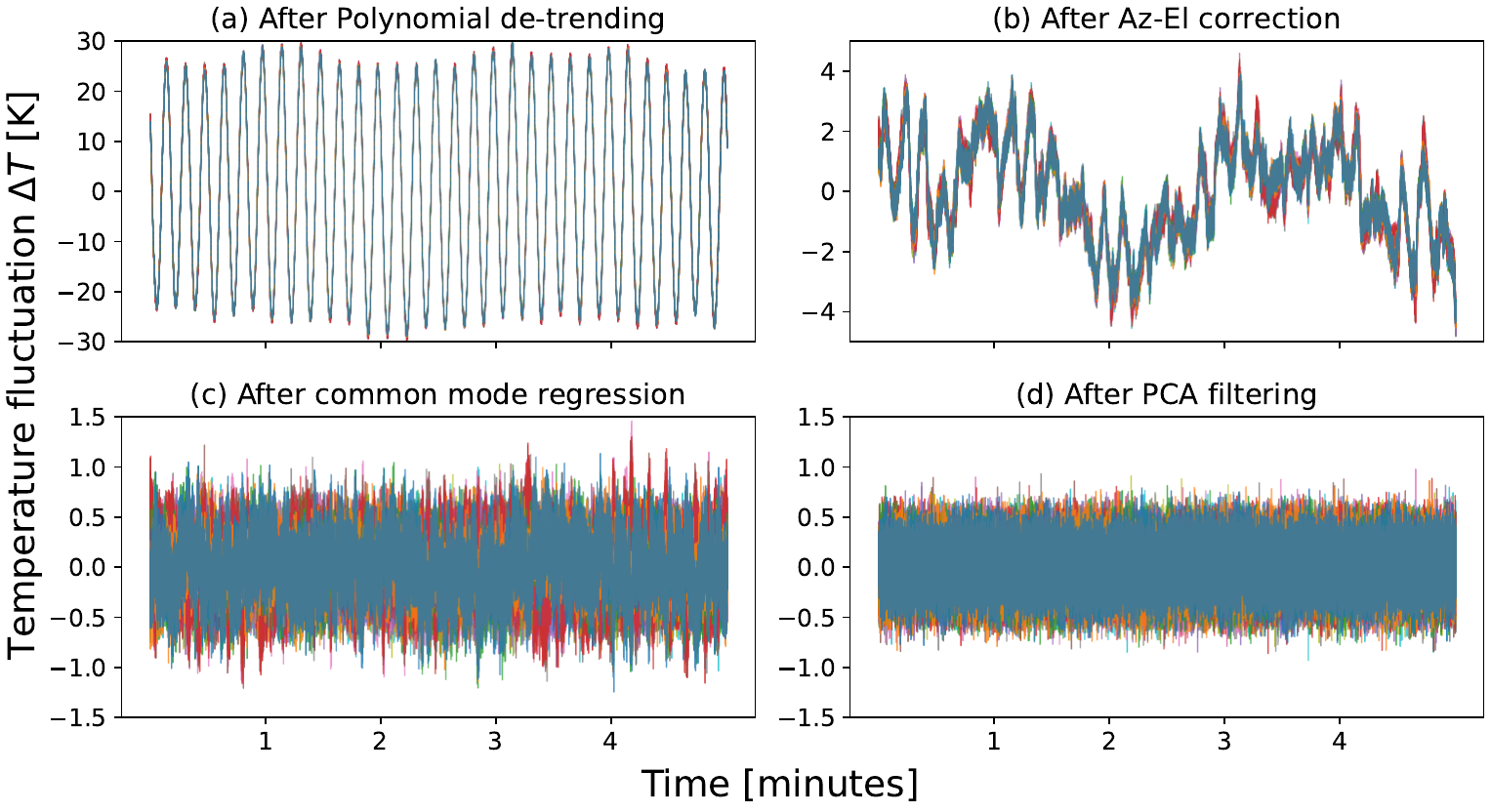}
  \caption{Same as \Fig{fig:2d_tod_clean}, but showing the detector timestreams for a selection of 10 detectors to illustrate the effect of each filtering stage. Panel (a) shows the timestreams after polynomial de-trending, panel (b) after correction for the Az-El scan motion, panel (c) after common-mode regression, and panel (d) after removing the leading three PCA components. Each timestream is normalised by its maximum absolute value to place all stages on a consistent scale.}
  \label{fig:1d_tod_clean}
\end{figure*}

\Figure{fig:1d_tod_clean} shows the effect of the successive filtering stages on the detector timestreams for a representative observation. After polynomial de-trending (panel a), the dominant scan-synchronous structure remains visible in the TOD. Correcting for the azimuth-elevation scan pattern (panel b) removes most of this large-scale variation, leaving residual correlated fluctuations between detectors. CM regression (panel c) further suppresses atmospheric fluctuations that are shared across the focal plane, substantially reducing the amplitude of the correlated signal. Finally, PCA filtering after removing the leading three principal components (panel d) removes the remaining dominant correlated modes, producing timestreams that are largely dominated by uncorrelated residual noise. 

\section{Maps for all Spectral Bins}
\begin{figure*}
\centering
  \includegraphics[width=17cm]{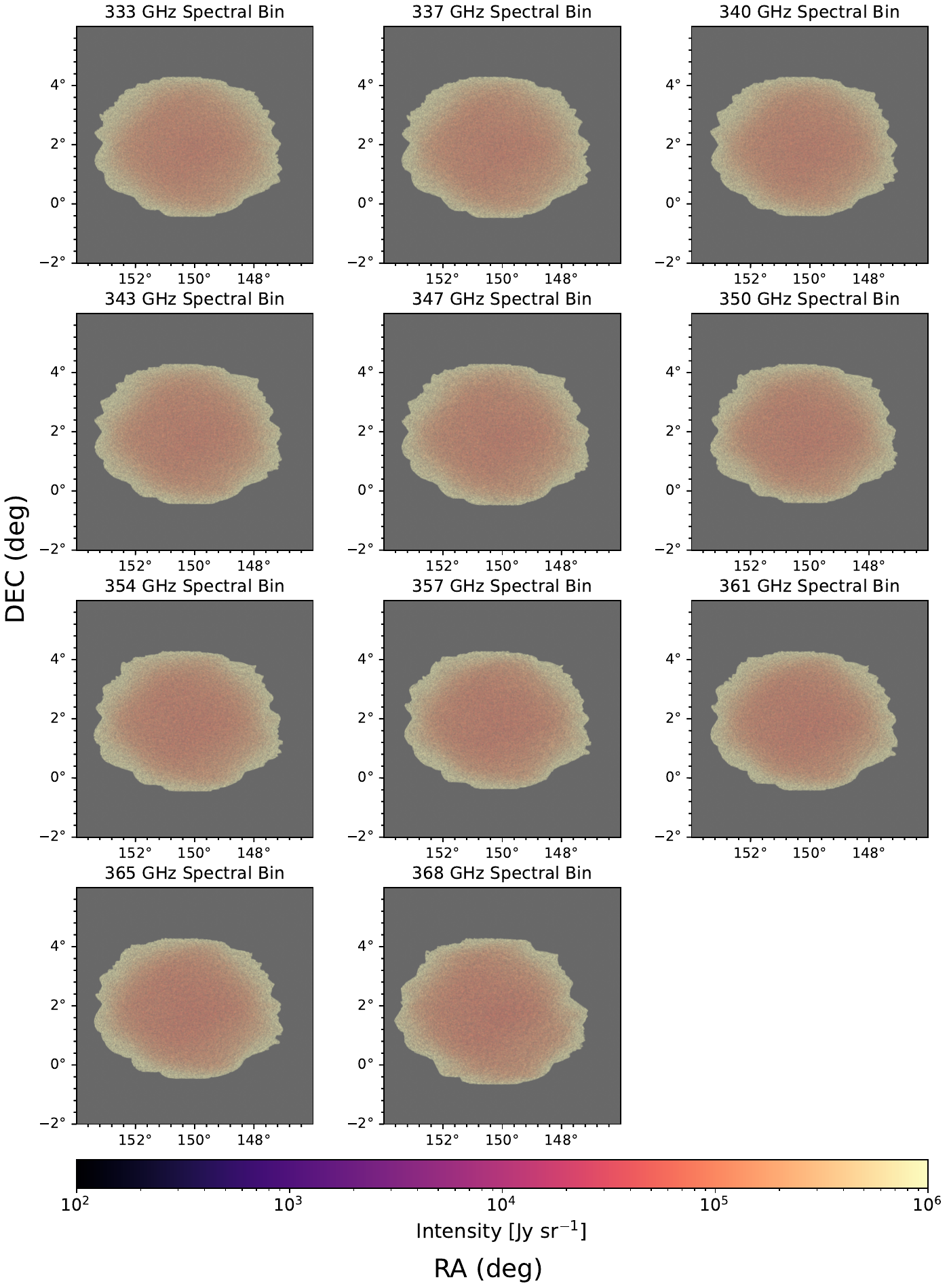}
  \caption{Mock maps for all the spectral bins across the $330 - 370$ GHz frequency range, after applying common mode regression and removing the leading 3 PCA components, showing the final filtered and binned outputs of the pipeline.}
  \label{fig:all_fbmaps_grid}
\end{figure*}

\Figure{fig:all_fbmaps_grid} shows the final co-added maps for all spectral bins spanning the $330 - 370$ GHz frequency range after applying the TOD filtering pipeline, map-making, and co-addition of all FPI step contributions associated with each spectral bin. These maps represent the final products used to construct the LIM data-cube for subsequent power-spectrum analysis.

Small differences in sky coverage are visible between spectral bins. This arises because each spectral bin receives contributions from a different subset of detectors and FPI step configurations, leading to slight variations in the effective footprint. Despite these differences, the central $4\,\mathrm{deg}^2$ region used for the LIM PS analysis has nearly uniform coverage across all spectral bins.

\end{appendix}

\end{document}